\documentclass{IEEEoj}

\usepackage{cite}
\usepackage{amsmath,amssymb,amsfonts}
\usepackage{algorithmic}
\usepackage{graphicx,color}
\usepackage{textcomp}
\usepackage{cite}
\usepackage{amsmath,amssymb,amsfonts}
\usepackage{bm}
\usepackage{algorithm}
\usepackage{algorithmic}
\usepackage{graphicx}
\usepackage{textcomp}
\usepackage{xcolor}
\usepackage{epsfig}
\usepackage{balance}
\usepackage[printonlyused]{acronym}
\usepackage{graphicx}
\usepackage{grffile}
\usepackage{stfloats}
\usepackage{type1cm}
\usepackage{float}
\newcommand{\CN}{\mathcal{CN}}

\newcommand{\diag}{\mathrm{diag}}

\acrodef{bs}[BS]{base station}
\acrodef{bss}[BSs]{base stations}
\acrodef{pl}[PL]{path loss}
\acrodef{ple}[PLE]{path loss exponent}
\acrodef{pdp}[PDP]{power delay profile}
\acrodef{cir}[CIR]{channel impulse response}
\acrodef{mpc}[MPC]{multi-path component}
\acrodef{mpcs}[MPCs]{multi-path components}

\def\BibTeX{{\rm B\kern-.05em{\sc i\kern-.025em b}\kern-.08em
    T\kern-.1667em\lower.7ex\hbox{E}\kern-.125emX}}
\AtBeginDocument{\definecolor{ojcolor}{cmyk}{0.93,0.59,0.15,0.02}}
\def\OJlogo{\vspace{-4pt}\hskip-4pt\includegraphics[height=18pt]{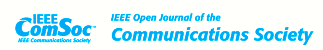}}

\begin{document}

\receiveddate{30 May}
\reviseddate{21 September 2026}
\accepteddate{}
\publisheddate{}
\currentdate{21 September 2026}
\doiinfo{}

\title{TRP--UE Bistatic Sensing for 6G\\ Performance Analysis and Insights}
\author{Necati Kagan Erkek\IEEEauthorrefmark{1,2} \IEEEmembership{(Student Member, IEEE)} and Marco Di Renzo \IEEEauthorrefmark{2,3}~\IEEEmembership{(Fellow,~IEEE)}, \\ Arman Shojaeifard \IEEEauthorrefmark{1}~\IEEEmembership{(Senior Member,~IEEE)} and Yasser Mestrah \IEEEauthorrefmark{1}, \\ Remun Koirala \IEEEauthorrefmark{1}, Mohammad Heggo\IEEEauthorrefmark{1}, Kunjan Shah\IEEEauthorrefmark{1} \IEEEmembership{(Member, IEEE)}}

\affil{InterDigital, Inc., London EC3A 3DH, United Kingdom}
\affil{King's College London, Centre for Telecommunications Research, London WC2R 2LS, United Kingdom}
\affil{CNRS and CentraleSup\'elec, Institute of Electronics and Digital Technologies (IETR), Avenue de la Boulaie, 35576 Cesson-S\'evign\'e, France}
\corresp{CORRESPONDING AUTHOR: Necati Kagan Erkek (e-mail: necati.erkek@kcl.ac.uk).}
\authornote{This work was supported by Marie Skłodowska-Curie Actions (MSCA) of EU and UK Research and Innovation (UKRI)}
\markboth{TRP--UE Bistatic Sensing for 6G: Performance Analysis and Insights}{Erkek \textit{et al.}}

\begin{abstract}
Integrated sensing and communication (ISAC) equips cellular networks with the dual capability of data transmission and environmental sensing by sharing spectrum, waveforms, and radio hardware.
While network-side sensing via transmission--reception points (TRPs) has been widely investigated, the use of user equipments (UEs) as sensing receivers introduces additional bistatic viewpoints and can improve target observability.
UE-assisted operation is nevertheless constrained by small UE apertures and by waveform processing, especially when data-bearing communication signals are reused for sensing.
The present paper investigates theoretical limits and operating conditions for UE-assisted passive target positioning, with emphasis on hybrid operation that fuses TRP--TRP and TRP--UE measurements.
A unified OFDM-based signal-and-noise model is developed to cover both positioning reference signals (PRS) and communication waveform reuse with modulation-dependent noise enhancement.
Based on this model, closed-form Fisher-information expressions and Cram\'er--Rao lower bounds are derived for delay and angular parameters and are mapped to positioning error bounds (PEBs) for two estimator classes: time--angle positioning based on TDOA+DoA and delay-only 3D ellipsoid positioning based on excess-delay/TDOA measurements. Monte-Carlo simulations across reference SNR, subcarrier spacing and occupied bandwidth, modulation order, radar cross section (RCS), sensing mode, and positioning algorithm identify regimes where UE participation and hybrid sensing provide significant gains and quantify the associated estimator-to-bound trade-offs.
\end{abstract}

\begin{IEEEkeywords}
ISAC, 5G-NR, 6G, target positioning, UE-assisted sensing, hybrid sensing, bistatic positioning, TDOA, DoA, communication waveform reuse, Positioning error bound (PEB). 
\end{IEEEkeywords}

\maketitle

\section{INTRODUCTION}
\vspace{4pt}

\IEEEPARstart{I}{ntegrated} sensing and communication (ISAC) is widely recognized as a key capability for beyond-5G and 6G systems, enabling simultaneous data transfer and environmental awareness through shared spectrum, waveforms, and radio hardware \cite{ref_erkek_vtc2026_hybrid, ref_erkek_isac2026_tracking, Liu2022Survey,Liu2020JCR,10077115,ZhangComST2022,LiuJSAC2022,PaulAccess2017,ChiriyathTCCN2017,ZhengSPM2019,FengChinaComm2020,MazahirFrontiers2021,WymeerschPIMRC2021}. 
The extension of dual-functional radar--communication concepts to cellular infrastructures has led to the perceptive mobile network paradigm, where distributed base stations provide sensing coverage in addition to communication services \cite{ZhangVTM2021,ZhangJSTSP2021,RahmanTAES2020,NiTSP2021,ZhangTVT2019Multibeam,RahmanComLett2021Clutter}. Standardization efforts further emphasize network sensing as a fundamental feature for future deployments \cite{3gppTR22837,ETSI_GR_ISC_001}. Thus, positioning of a passive sensing target (ST) constitutes a core ISAC functionality, relying on the estimation of delay, Doppler, and angular parameters, followed by position inference. In cellular deployments, sensing may be performed solely by TRPs, through UE-assisted configurations where UEs act as sensing receivers, or via hybrid fusion of TRP--TRP and TRP--UE measurements. UE participation increases spatial diversity and improves geometric conditions; however, limitations arise due to smaller UE array apertures, reduced sensing SNR, and waveform-dependent processing losses.

\subsection{Existing Studies}
\vspace{4pt}
Information-theoretic limits for parameter estimation in cooperative MIMO-OFDM ISAC networks are derived in \cite{Pucci2025Fisher}, covering monostatic, bistatic, and cooperative configurations under communication waveform reuse and uniform linear array (ULA) assumptions. 
Pilot-aided sensing based on 5G NR positioning reference signals (PRS) is investigated in \cite{WeiPRS2022}, where delay and velocity estimation performance and corresponding Cram\'er--Rao lower bounds are obtained. However, communication waveform reuse and angular estimation aspects are not addressed therein. Foundational dual-function radar--communication frameworks and waveform-fusion concepts are surveyed in \cite{SturmWiesbeck2011,Hassanien2019DFRC}, while system-level signal processing perspectives for ISAC are summarized in \cite{ZhangJSTSP2021,Liu2022Survey, SchmidtTAP1986, SalehJSAC1987, RappaportAccess2013, SturmWiesbeck2011, Hassanien2019DFRC, WaxTASSP1985}.

More recent studies investigate cooperative ISAC sensing from multiple perspectives, including cooperative positioning algorithms, network optimization, passive sensing with communication signals, and multistatic sensing architectures \cite{10912515,10694260,10769538,10787076,10736660,11203895,11274663}. A broader system-level discussion addressing synchronization, beam management, and deployment challenges is provided in \cite{11173647}. In addition, several works emphasize the impact of network densification and distributed processing on sensing accuracy, highlighting the importance of spatial diversity and coordination among nodes. The role of hardware constraints, such as limited-resolution ADCs and practical RF impairments, has also been recognized as a key factor affecting achievable sensing performance in realistic deployments. Furthermore, recent contributions point out the sensitivity of cooperative sensing performance to waveform design parameters, such as subcarrier spacing, bandwidth allocation, and modulation schemes, which directly influence estimation accuracy and robustness. Despite these contributions, a unified treatment of hybrid TRP--TRP and TRP--UE sensing with realistic array configurations, waveform-dependent effects, and estimator-aligned performance bounds remains limited.

\subsection{Contribution and Improvement}
\vspace{4pt}
This paper addresses the above gap by developing a unified theoretical and simulation-based framework for UE-assisted passive target positioning in cellular ISAC networks, with particular emphasis on hybrid sensing that fuses TRP--TRP and TRP--UE measurements. The proposed framework is grounded in practical OFDM signaling and realistic URA configurations at both infrastructure nodes and UEs, thereby ensuring that the analysis remains closely aligned with implementable receiver structures and realistic deployment assumptions. Beyond establishing fundamental performance limits, the paper also seeks to clarify the conditions under which UE assistance enhances positioning robustness, as opposed to cases in which performance remains constrained by aperture, geometry, or waveform-related losses. The main contributions of this work are summarized as follows:

\begin{itemize}
    \item A unified OFDM-based signal, channel, and noise model is established to cover both PRS-based sensing and communication waveform reuse. The model explicitly captures the modulation-dependent noise enhancement that arises after data-symbol reconstruction, thereby enabling a fair comparison between pilot-based and reuse-based sensing.
    
    \item Closed-form Fisher information expressions and the associated Cram\'er--Rao lower bounds are derived for delay and angular parameters under practical URA deployments at both TRPs and UEs. This analysis makes explicit the interplay between occupied bandwidth and delay resolution, as well as that between array aperture and angular resolution, thereby revealing the principal scaling laws governing positioning accuracy across heterogeneous sensing links.
    
    \item Estimator-aligned positioning error bounds (PEBs) are developed for two representative positioning pipelines: a time--angle approach based on TDOA+DoA and a delay-only 3D ellipsoid multilateration approach based on excess-delay/TDOA measurements. These bounds provide a rigorous benchmark for assessing the extent to which the corresponding Monte-Carlo estimators operate near their theoretical limits.
    
    \item Three sensing modes---TRP--TRP, TRP--UE, and Hybrid---are evaluated in a realistic indoor-factory deployment comprising multiple TRPs, UEs, and passive targets. The analysis systematically examines the impact of SNR, subcarrier spacing and occupied bandwidth, modulation order, target radar cross section (RCS), and sensing-pair diversity on positioning accuracy and robustness.
    
    \item The operating regimes in which UE participation and hybrid fusion provide the greatest gains are identified, and the dependence of these gains on the balance between viewpoint diversity and receiver angular resolution is characterized. In particular, the results quantify the gap between delay-only multilateration and its corresponding PEB, as well as the cases in which time--angle positioning remains limited by angular estimation accuracy.
\end{itemize}

The remainder of this paper is organized as follows: Section~\ref{sec:signal_noise_model} establishes the signal and noise models, Section~\ref{sec:boundary_analysis} derives the Cram'er--Rao and positioning error bounds, and Section~\ref{sec:postioning} details the positioning algorithms. Finally, Section~\ref{sec:sim_env} outlines the simulation setup and parameters, Section~\ref{sec:sim_results} evaluates the numerical results, and Section~\ref{sec:conclusion} concludes the paper.

\section{Signal, Noise and Channel Model}
\label{sec:signal_noise_model}
\vspace{8pt}
\subsection{Sensing Signal Models}
\label{subsec:sensing_signal_models}

Two OFDM-based sensing modes are considered. The first one is \emph{PRS-based sensing}, where the 5G positioning reference signal (PRS) symbols are known at the sensing receiver. The second one is \emph{communication waveform reuse}, where a data-bearing communication signal is reused for sensing once the receiver has access to (or reconstructs) the transmitted data symbols.
After specifying the signal generation/transmission model, antenna, channel, and received-signal models are common to PRS-based sensing and communication waveform reuse.

\vspace{-12pt}
\subsubsection{PRS-Based Sensing Signal Model}
\label{subsubsec:prs_sensing_model}
\vspace{4pt}
In 5G NR, the PRS sequence associated with OFDM symbol $\ell$ is generated as \cite{3gppTS38211,WeiPRS2022}
\begin{equation}
r_{\ell}[n]
=
\frac{1}{\sqrt{2}}\big(1-2c_{\ell}(2n)\big)
+j\frac{1}{\sqrt{2}}\big(1-2c_{\ell}(2n+1)\big)
\label{eq:prs_sequence}
\end{equation}
$n=0,\ldots,K_{\ell}-1$, where $n$ is the PRS-sequence index and $c_{\ell}(i)$ is the Gold sequence initialized for symbol $\ell$ according to the PRS identity and slot configuration \cite{3gppTS38211}. Let $\ell_0$ be the configured starting symbol, let $N_{\mathrm{sym}}^{\mathrm{PRS}}$ be the number of consecutive PRS symbols, and define $\mathcal L_{\mathrm{PRS}}\triangleq\{\ell_0,\ldots,\ell_0+N_{\mathrm{sym}}^{\mathrm{PRS}}-1\}$. Furthermore, let $N_{\mathrm{RB}}^{\mathrm{PRS}}$ be the number of allocated PRBs, $N_{\mathrm{RB}}^{\mathrm{start}}$ the RB offset within the carrier grid, $K_{\mathrm{comb}}$ the comb size, $k_{\mathrm{RE}}^{\mathrm{off}}$ the configured RE offset, and $\kappa_{\ell}$ the symbol-dependent comb offset specified by the NR PRS mapping. The $n$th seq. sample in symbol $\ell$ is mapped to carrier-grid subcarrier
\begin{equation}
k_{\ell,n}
=
12N_{\mathrm{RB}}^{\mathrm{start}}
+
\operatorname{mod}\!\left(k_{\mathrm{RE}}^{\mathrm{off}}+\kappa_{\ell},K_{\mathrm{comb}}\right)
+nK_{\mathrm{comb}}
\label{eq:prs_comb_freq}
\end{equation}
\begin{equation}
K_{\ell}=\frac{12N_{\mathrm{RB}}^{\mathrm{PRS}}}{K_{\mathrm{comb}}}
\label{eq:prs_comb_freq2}
\end{equation}
Accordingly, the mapped resource-grid symbol is
\begin{equation}
s_{\mathrm{PRS}}[k,\ell]
=
\begin{cases}
\rho_{\mathrm{PRS}}\,r_{\ell}[n], & k=k_{\ell,n},\\
0, & \text{otherwise},
\end{cases}
\label{eq:prs_re_mapping}
\end{equation}
where $\rho_{\mathrm{PRS}}$ is the PRS amplitude-normalization factor; $\rho_{\mathrm{PRS}}=1$ is used in this work, so every active PRS symbol has unit modulus. PRS resource-element set is
\begin{equation}
\mathcal A_{\mathrm{PRS}}
\triangleq
\left\{(k_{\ell,n},\ell):\ell\in\mathcal L_{\mathrm{PRS}},\ n=0,\ldots,K_{\ell}-1\right\}.
\label{eq:prs_active_set}
\end{equation}
Adjacent active PRS tones within one OFDM symbol are separated by $K_{\mathrm{comb}}\Delta f$, while the residue class can change with $\ell$ through $\kappa_{\ell}$. If $t_{\ell}$ denotes reference time of OFDM symbol $\ell$, the transmitted PRS-mapped signal:
\begin{equation}
x_{\mathrm{PRS}}(t)
=
\sum_{(k,\ell)}
 s_{\mathrm{PRS}}[k,\ell]\,
 e^{j2\pi f_k(t-t_{\ell})}\,
\mathrm{rect}\!\left(\frac{t-t_{\ell}}{T_{s,\ell}}\right)
\label{eq:prs_tx_time}
\end{equation}
where $f_k=(k-k_{\mathrm{ref}})\Delta f$ baseband subcarrier frequency and $T_{s,\ell}$ includes the useful OFDM duration and the symbol-dependent cyclic prefix. Equations~\eqref{eq:prs_sequence}--\eqref{eq:prs_active_set} explicitly connect the Gold-sequence sample, resource element, OFDM-symbol index, comb offset, and normalization.

\subsubsection{Communication Waveform Reuse Sensing Model}
\label{subsubsec:data_ofdm_reuse_model}
\vspace{4pt}
Let $d(k,m)\in\mathbb{C}$ denote the communication data symbol transmitted on subcarrier $k$ and OFDM symbol $m$ (e.g., $M$-QAM).
In communication waveform reuse, these data symbols are treated as known at the sensing processor once decoding and re-encoding (or other reconstruction) are performed. Let $\mathcal{K}$ denote the set of used subcarriers for the data signal (typically $|\mathcal{K}|=N$ for an $N$-point IFFT/FFT).
The corresponding continuous-time transmit signal is
\begin{equation}
x_{\mathrm{D}}(t)=\sum_{m=0}^{M-1}\sum_{k\in\mathcal{K}} d(k,m)\,
e^{j2\pi f_k t}\,
\mathrm{rect}\!\left(\frac{t-mT_s}{T_s}\right)
\label{eq:data_tx_time}
\end{equation}

For the communication waveform, define $\mathcal A_{\mathrm D}\triangleq\{(k,m):d[k,m]\neq0\}$. From this point onward, $\mathcal A$ denotes the active resource-element set of the selected waveform, i.e., $\mathcal A=\mathcal A_{\mathrm{PRS}}$ or $\mathcal A=\mathcal A_{\mathrm D}$, and $N_{\mathcal A}\triangleq |\mathcal A|$. The symbol $x[k,m]$ equals $s_{\mathrm{PRS}}[k,m]$ for PRS-based sensing and $d[k,m]$ for communication waveform reuse. The corresponding baseband frequency $f_k$ and OFDM-symbol reference time $t_m$ are retained for every pair $(k,m)\in\mathcal A$. All subsequent antenna, channel, receiver, and Fisher-information models use this same active set; in particular, the PRS analysis does not replace its comb allocation by a contiguous-tone surrogate.

\subsection{URA Antenna and Steering Vectors}
\label{subsec:ura_steering_vectors}

Uniform rectangular arrays (URAs) are considered at the transmitter and receiver. The half-wavelength spacing is adopted as $d_{T,x}=d_{T,y}=d_{R,x}=d_{R,y}=\lambda/2$, where $\lambda=c/f_c$. The TRP is equipped with an $8\times 8$ URA and the UE is equipped with a $2\times 2$ URA. The same URA model applies if the TRP is used as the sensing receiver.

\paragraph{Direction cosines}
Let $(\varphi,\vartheta)$ denote azimuth and elevation angles, and define the direction cosines as
\begin{equation}
u_x(\varphi,\vartheta) \triangleq \cos\vartheta\cos\varphi,
\qquad
u_y(\varphi,\vartheta) \triangleq \cos\vartheta\sin\varphi
\label{eq:dir_cosines}
\end{equation}

\paragraph{Transmit steering vector}
For a Tx URA of size $N_{T,x}\times N_{T,y}$, the 2D steering vector:
\begin{equation}
\bm a(\varphi_T,\vartheta_T)
=
\bm a_y(\varphi_T,\vartheta_T)\otimes \bm a_x(\varphi_T,\vartheta_T)
\label{eq:tx_steering}
\end{equation}
where $\otimes$ is the Kronecker product and its components:
\begin{equation}
\begin{aligned}
\bm a_x{_{(\varphi_T,\vartheta_T)}}
&=
\big[
e^{-j2\pi\frac{d_{T,x}}{\lambda}\left(p-\frac{N_{T,x}-1}{2}\right)u_x{_{(\varphi_T,\vartheta_T)}}}
\big]_{p=0}^{N_{T,x}-1}
\\
\bm a_y{_{(\varphi_T,\vartheta_T)}}
&=
\big[
e^{-j2\pi\frac{d_{T,y}}{\lambda}\left(q-\frac{N_{T,y}-1}{2}\right)u_y{_{(\varphi_T,\vartheta_T)}}}
\big]_{q=0}^{N_{T,y}-1}
\end{aligned}
\end{equation}
\vspace{-1cm}
\paragraph{Receive steering vector.}
For a receive URA of size $N_{R,x}\times N_{R,y}$, the 2D steering vector is defined below:
\begin{equation}
\bm b(\varphi_R,\vartheta_R)
=
\bm b_y(\varphi_R,\vartheta_R)\otimes \bm b_x(\varphi_R,\vartheta_R)
\label{eq:rx_steering}
\end{equation}
with the following $x$ and $y$ components below:
\begin{equation}
\begin{aligned}
\bm b_x(\varphi_R,\vartheta_R)
&=
\big[
e^{-j2\pi\frac{d_{R,x}}{\lambda}\left(m-\frac{N_{R,x}-1}{2}\right)u_x(\varphi_R,\vartheta_R)}
\big]_{m=0}^{N_{R,x}-1}
\\
\bm b_y(\varphi_R,\vartheta_R)
&=
\big[
e^{-j2\pi\frac{d_{R,y}}{\lambda}\left(n-\frac{N_{R,y}-1}{2}\right)u_y(\varphi_R,\vartheta_R)}
\big]_{n=0}^{N_{R,y}-1}
\end{aligned}
\end{equation}

\vspace{-8pt}
\subsection{Channel Model}
\label{subsec:channel_model}
\vspace{4pt}
A single-target (single-scatterer) bistatic channel model is adopted.
Let $d_T$ and $d_R$ denote the Tx-to-ST and ST-to-receiver distances, respectively~\cite{10012846}, and let $d_{TR}$ denote the direct Tx-to-receiver distance. The active data/pilot symbols satisfy $\mathbb{E}\{|x[k,m]|^2\}=1$. We denote by $P_{\mathrm{RE}}$ the average transmit power allocated to one active subcarrier; if $K_{\mathrm a}$ subcarriers are simultaneously active with the same allocation, the corresponding total OFDM-symbol transmit power is $P_T=K_{\mathrm a}P_{\mathrm{RE}}$. Let $G_T$ and $G_R$ denote scalar transmit and receive propagation/element gains and let $\sigma_{\mathrm{RCS}}$ denote the target RCS in m$^2$. The modeled transmit-array factor $\beta$ and receive-array response $\bm b$ are excluded from $G_T$ and $G_R$ to avoid counting array gain twice. The bistatic power gain:
\begin{equation}
G_{\mathrm{bi}}
=
\frac{G_TG_R\lambda^2\sigma_{\mathrm{RCS}}}
{(4\pi)^3d_T^2d_R^2}
\label{eq:bistatic_power_gain}
\end{equation}
For the direct baseline component, free-space power gain:
\begin{equation}
G_{\mathrm{LOS}}
=
G_TG_R\left(\frac{\lambda}{4\pi d_{TR}}\right)^2
\label{eq:los_power_gain}
\end{equation}
Both $G_{\mathrm{bi}}$ and $G_{\mathrm{LOS}}$ are power ratios. Therefore, the complex baseband coefficients scale with their square roots. In particular, letting $\tau$ and $f_D$ denote the bistatic propagation delay and Doppler shift, frequency-domain MIMO channel after CP removal and FFT:
\begin{equation}
\bm H_{[k,m]}
=
\sqrt{G_{\mathrm{bi}}}\,e^{j\phi_0}
e^{j2\pi t_m f_D}
e^{-j2\pi f_k\tau}
\bm b_{(\varphi_R,\vartheta_R)}\bm a^H_{(\varphi_T,\vartheta_T)}
\label{eq:channel_rank1}
\end{equation}
where $f_k$ and $t_m$ are the baseband frequency and reference time associated with active resource element $(k,m)$, $\phi_0$ is an unknown constant phase, and $(\varphi_T,\vartheta_T)$ and $(\varphi_R,\vartheta_R)$ are the departure and arrival angles at the transmitter and receiver.

\subsection{Received Signal}

\vspace{4pt}
Assume single-stream transmission with a deterministic transmit-excitation vector $\bm w_T\in\mathbb{C}^{N_T\times 1}$, $\|\bm w_T\|^2=1$. No specific beam-steering or beamforming optimization is required in the derivations below; $\bm w_T$ captures a fixed excitation across the transmit URA. The transmitted antenna-domain vector is
$\bm x_T[k,m]=\sqrt{P_{\mathrm{RE}}}\bm w_Tx[k,m]$,
and the received vector is
\vspace{-4pt}
\begin{align}
\bm y[k,m]
&=\bm H[k,m]\bm x_T[k,m]+\bm n[k,m]
\label{eq:rx_signal_general}
\end{align}
Substituting \eqref{eq:channel_rank1} and $\bm x_T[k,m]$ into \eqref{eq:rx_signal_general} gives
\begin{align}
\bm y[k,m]
&=
\sqrt{P_{\mathrm{RE}}G_{\mathrm{bi}}}\,e^{j\phi_0}
e^{j2\pi t_m f_D}e^{-j2\pi f_k\tau}
\bm b(\varphi_R,\vartheta_R)
\nonumber\\
&\quad\times
\underbrace{\bm a^H(\varphi_T,\vartheta_T)\bm w_T}_{\triangleq\,\beta(\varphi_T,\vartheta_T)}
x[k,m]+\bm n[k,m]
\label{eq:rx_signal_beamformed}
\end{align}
Here, $\bm n[k,m]\sim\CN(\bm 0,\sigma_{N,\mathrm{RE}}^2\bm I_{N_R})$ is spatially white AWGN at one FFT output. With one-sided equivalent noise bandwidth $B_{\mathrm{RE}}=\Delta f$, $\sigma_{N,\mathrm{RE}}^2=k_{\mathrm B}T_0F\Delta f$, where $k_{\mathrm B}$ is Boltzmann's constant, $T_0$ is the receiver-noise temperature, and $F$ is the linear noise factor. If OFDM symbol $m$ contains $K_{\mathrm a,m}$ active tones, then $P_{T,m}=K_{\mathrm a,m}P_{\mathrm{RE}}$ and the summed FFT-bin noise power is $k_{\mathrm B}T_0F K_{\mathrm a,m}\Delta f$. This per-symbol notation remains valid for the sparse PRS comb and avoids confusing active-tone count with frequency aperture.

\paragraph{Common normalized observation}
For both PRS-based sensing and communication waveform reuse, it is assumed that $x[k,m]$ is available to the sensing processor. For PRS, $x[k,m]$ is a pilot and is known by design; for the reuse case, $x[k,m]$ is obtained via data reconstruction.
The normalized observation as:
\begin{equation}
\bm z[k,m]= \frac{\bm y[k,m]}{x[k,m]},
\qquad (k,m)\in\mathcal A
\label{eq:zkm_normalized}
\end{equation}

\begin{equation}
\bm z[k,m]=
\gamma\,
e^{j2\pi t_m f_D}\,
e^{-j2\pi f_k\tau}
\bm b(\varphi_R,\vartheta_R)
+\bm w[k,m]
\label{eq:zkm_normalized2}
\end{equation}

where $\gamma\triangleq\sqrt{P_{\mathrm{RE}}G_{\mathrm{bi}}}\,e^{j\phi_0}\beta(\varphi_T,\vartheta_T)$ and $\bm w[k,m]=\bm n[k,m]/x[k,m]$. The division is applied only on the set $\mathcal{A}$ of active resource elements for which $x[k,m]\neq 0$. Conditioned on the known symbol realization, the normalized noise is circularly symmetric Gaussian but generally heteroscedastic:
\begin{equation}
\bm w[k,m]\mid x[k,m]
\sim
\CN\!\left(
\bm 0,
\frac{\sigma_{N,\mathrm{RE}}^2}{|x[k,m]|^2}\bm I_{N_R}
\right), (k,m)\in\mathcal{A}.
\label{eq:conditional_normalized_noise}
\end{equation}
For the adopted \emph{unweighted reciprocal-normalization receiver}, a frame-wise equivalent average variance as:
\begin{equation}
\overline{\sigma}_{e,N}^{2}
\triangleq
\frac{\sigma_{N,\mathrm{RE}}^2}{|\mathcal{A}|}
\sum_{(k,m)\in\mathcal{A}}
\frac{1}{|x[k,m]|^2}
\triangleq
\sigma_{N,\mathrm{RE}}^2\eta_{\mathcal{A}}
\label{eq:effective_noise2}
\end{equation}

\begin{equation}
\eta_{\mathcal{A}}\approx
\mathbb{E}\!\left\{\frac{1}{|x[k,m]|^2}\right\}
\label{eq:effective_noise22}
\end{equation}

For constant-modulus PRS, BPSK, and QPSK symbols, $\eta_{\mathcal{A}}=1$, and the conditional and equivalent noise models coincide. For non-constant-modulus QAM, $\eta_{\mathcal{A}}>1$ and the individual variances remain symbol dependent. Under unit-average-power square-QAM normalization, the ensemble values are approximately $1.89$ for 16QAM and $2.69$ for 64QAM, corresponding to average receiver-output shifts of $2.76$~dB and $4.29$~dB, respectively. These shifts characterize the adopted unweighted reciprocal receiver; they are not fundamental limits of communication-waveform reuse, since symbol-aware weighting or a matched-likelihood receiver can exploit the known amplitudes differently.

\paragraph{Remark on the effective sensing SNR}
For the normalized observation model in \eqref{eq:zkm_normalized}, each entry of $\bm b(\varphi,\vartheta)$ has unit modulus. Since
$|\gamma|^2=P_{\mathrm{RE}}G_{\mathrm{bi}}|\beta(\varphi_T,\vartheta_T)|^2$,
the target-echo power at one receive element is $P_{\mathrm{RE}}G_{\mathrm{bi}}|\beta|^2$, while coherent collection across the URA contributes the separate factor $\|\bm b\|^2=N_R$, where $N_R=N_{R,x}N_{R,y}$. Accordingly, the per-element target-echo SNR before reciprocal normalization is:
\begin{equation}
\mathrm{SNR}_{\mathrm{pre}}
\triangleq
\frac{P_{\mathrm{RE}}G_{\mathrm{bi}}|\beta|^2\,\mathbb{E}\!\left\{|x[k,m]|^2\right\}}
{\sigma_{N,\mathrm{RE}}^2}
=
\frac{P_{\mathrm{RE}}G_{\mathrm{bi}}|\beta|^2}{\sigma_{N,\mathrm{RE}}^2}.
\label{eq:snr_pre}
\end{equation}
For uniform loading over $K_{\mathrm a}$ tones, $P_T=K_{\mathrm a}P_{\mathrm{RE}}$ and $B=K_{\mathrm a}\Delta f$, so the same quantity can equivalently be written as
$\mathrm{SNR}_{\mathrm{pre}}=P_TG_{\mathrm{bi}}|\beta|^2/(k_{\mathrm B}T_0FB)$.
direct-path reference is obtained from transmit and noise power:
\begin{equation}
\mathrm{SNR}_{\mathrm{LOS,pre}}
=
\frac{P_{\mathrm{RE}}G_{\mathrm{LOS}}|\beta_{\mathrm{LOS}}|^2}{\sigma_{N,\mathrm{RE}}^2}
\label{eq:snr_los_pre}
\end{equation}
For the equivalent average-noise representation of the unweighted reciprocal receiver,
\begin{equation}
\mathrm{SNR}_{\mathrm{eff}}
\triangleq
\frac{P_{\mathrm{RE}}G_{\mathrm{bi}}|\beta|^2}{\overline{\sigma}_{e,N}^{2}}
=
\frac{P_{\mathrm{RE}}G_{\mathrm{bi}}|\beta|^2}{\eta_{\mathcal{A}}\sigma_{N,\mathrm{RE}}^2}
=
\frac{\mathrm{SNR}_{\mathrm{pre}}}{\eta_{\mathcal{A}}}.
\label{eq:snr_eff}
\end{equation}
For constant-modulus signaling, this expression is exact. For non-constant-modulus QAM, it is a moment-matched, receiver-aligned approximation to the exact heteroscedastic model in \eqref{eq:conditional_normalized_noise}. Accordingly, the common equivalent-FIM factor is written as
\begin{equation}
\frac{2|\gamma|^2}{\overline{\sigma}_{e,N}^{2}}
=
2\,\mathrm{SNR}_{\mathrm{eff}},
\qquad
\frac{2|\gamma|^2N_R}{\overline{\sigma}_{e,N}^{2}}
=
2N_R\,\mathrm{SNR}_{\mathrm{eff}}.
\label{eq:fim_snr_factor}
\end{equation}
Definition does not require a particular Tx beamformer: any deterministic illumination factor is represented explicitly through $|\beta|^2$, while $G_{\mathrm{bi}}$ remains a power gain.

\section{Bound Analysis}
\label{sec:boundary_analysis}
\vspace{6pt}
\subsection{CRLB Derivation by Likelihood Estimation}
\label{subsec:crlb_derivation}

\vspace{4pt}

Following the standard likelihood/FIM framework \cite{WeiPRS2022} specialized for URA observations in \eqref{eq:zkm_normalized}, the CRLBs are derived by eliminating the complex target-reflection coefficient via a Schur complement. These bounds are exact for constant-modulus signals and serve as equivalent-noise benchmarks for 16QAM and 64QAM via \eqref{eq:effective_noise2}, while Monte Carlo methods retain the exact symbol-by-symbol variances in \eqref{eq:conditional_normalized_noise}. Additionally, the single-path delay bound serves as an isolated-path reference, distinct from the excess-delay PEB in Section~\ref{sec:postioning} which employs a joint two-path FIM to form the LOS--target delay covariance.
\vspace{-16pt}
\subsubsection{Open form of the received signal}

From \eqref{eq:zkm_normalized}, each active observation satisfies:
\begin{equation}
\bm z[k,m]=\bm \mu[k,m](\bm\theta)+\bm w[k,m],
\qquad (k,m)\in\mathcal A,
\label{eq:zkm_mu_noise}
\end{equation}
\begin{equation}
\bm \mu[k,m](\bm\theta)
=
\gamma\,
e^{j2\pi t_m f_D}\,
e^{-j2\pi f_k\tau}\,
\bm b(\varphi,\vartheta).
\label{eq:mu_km}
\end{equation}
Write the unknown complex target-reflection coefficient as $\gamma=\gamma_{\mathrm R}+j\gamma_{\mathrm I}$. Parameters of interest are $\bm\theta_{\mathrm s}
\triangleq
[\tau,\ f_D,\ \varphi,\ \vartheta]^T,
$ and the complete real-valued parameter vector is $
\bm\theta
\triangleq
[\bm\theta_{\mathrm s}^{T},\ \gamma_{\mathrm R},\ \gamma_{\mathrm I}]^T.
$
Neither echo of amplitude nor its initial phase assumed as a known parameter.
\vspace{-8pt}
\subsubsection{Open form likelihood and log-likelihood function}
For constant-modulus signaling, $\overline{\sigma}_{e,N}^{2}=\sigma_{N,\mathrm{RE}}^2$ and the normalized active samples are identically distributed. For non-constant-modulus QAM, following homogeneous-noise likelihood is the receiver-aligned average-noise model associated with the adopted unweighted reciprocal processor. With $\mathcal{Z}\triangleq\{\bm z[k,m]:(k,m)\in\mathcal A\}$,
\begin{equation}
p(\mathcal{Z};\bm\theta)
=
\prod_{(k,m)\in\mathcal A}
\frac{1}{(\pi\overline{\sigma}_{e,N}^{2})^{N_R}}
e^{(
-\frac{\left\|\bm z[k,m]-\bm \mu[k,m](\bm\theta)\right\|^2}
{\overline{\sigma}_{e,N}^{2}}
)}
\label{eq:likelihood}
\end{equation}
The corresponding log-likelihood is
\begin{align}
\ell(\bm\theta)
&=
- N_{\mathcal A}N_R\log(\pi\overline{\sigma}_{e,N}^{2})
\nonumber\\
&\quad
-\frac{1}{\overline{\sigma}_{e,N}^{2}}
\sum_{(k,m)\in\mathcal A}
\left\|\bm z[k,m]-\bm \mu[k,m](\bm\theta)\right\|^2
\label{eq:loglik}
\end{align}

\subsubsection{Open form Fisher information matrix (FIM)}
For real-valued parameters $\theta_i$ and known $\overline{\sigma}_{e,N}^{2}$, FIM:
\begin{equation}
\left[\bm F(\bm\theta)\right]_{i,j}
=
\frac{2}{\overline{\sigma}_{e,N}^{2}}
\Re\!\left\{
\sum_{(k,m)\in\mathcal A}
\frac{\partial \bm \mu^H[k,m]}{\partial \theta_i}
\frac{\partial \bm \mu[k,m]}{\partial \theta_j}
\right\}.
\label{eq:fim_general}
\end{equation}

\paragraph{Centered active frequency--time coordinates.}
The centering must be performed over the \emph{actual} active RE set rather than over fictitious contiguous indices.
\begin{equation}
\begin{aligned}
\overline f_{\mathcal A}
&\triangleq \frac{1}{N_{\mathcal A}}\sum_{(k,m)\in\mathcal A}f_k,
&
\widetilde f_{k,m}
&\triangleq f_k-\overline f_{\mathcal A},\\
\overline t_{\mathcal A}
&\triangleq \frac{1}{N_{\mathcal A}}\sum_{(k,m)\in\mathcal A}t_m,
&
\widetilde t_{k,m}
&\triangleq t_m-\overline t_{\mathcal A}.
\end{aligned}
\label{eq:active_coordinate_centering}
\end{equation}
Hence, $\sum_{\mathcal A}\widetilde f_{k,m}=\sum_{\mathcal A}\widetilde t_{k,m}=0$. The common phase generated by centering is absorbed into $\gamma$.

\paragraph{Derivatives of the mean.}
With the centered coordinates in \eqref{eq:active_coordinate_centering},
\begin{align}
\frac{\partial \bm \mu_{[k,m]}}{\partial \tau}
&=-j2\pi\widetilde f_{k,m}\bm \mu_{[k,m]},
&
\frac{\partial \bm \mu_{[k,m]}}{\partial f_D}
&=j2\pi\widetilde t_{k,m}\bm \mu_{[k,m]}
\label{eq:deriv_tau_fd}
\end{align}
\begin{align}
\frac{\partial \bm \mu_{[k,m]}}{\partial \varphi}
&=\gamma q_{k,m}\frac{\partial\bm b(\varphi,\vartheta)}{\partial\varphi},
&
\frac{\partial \bm \mu_{[k,m]}}{\partial \vartheta}
&=\gamma q_{k,m}\frac{\partial\bm b(\varphi,\vartheta)}{\partial\vartheta}
\label{eq:deriv_angles}
\end{align}
where $q_{k,m}\triangleq e^{j2\pi\widetilde t_{k,m}f_D}e^{-j2\pi\widetilde f_{k,m}\tau}$. The nuisance-gain derivatives are:
\begin{align}
\frac{\partial \bm\mu[k,m]}{\partial \gamma_{\mathrm R}}
&=q_{k,m}\bm b(\varphi,\vartheta),
&
\frac{\partial \bm\mu[k,m]}{\partial \gamma_{\mathrm I}}
&=j q_{k,m}\bm b(\varphi,\vartheta)
\label{eq:deriv_complex_gain}
\end{align}

\paragraph{Equivalent information after eliminating complex gain.}
Partition the augmented FIM according to $\bm\theta_{\mathrm s}$ and $\bm\gamma_{\mathrm r}\triangleq[\gamma_{\mathrm R},\gamma_{\mathrm I}]^T$ as
\begin{equation}
\bm F(\bm\theta)
=
\begin{bmatrix}
\bm F_{\mathrm{ss}} & \bm F_{\mathrm{s}\gamma}\\
\bm F_{\gamma\mathrm{s}} & \bm F_{\gamma\gamma}
\end{bmatrix},
\qquad
\bm F_{\gamma\gamma}
=
\frac{2N_{\mathcal A}N_R}{\overline{\sigma}_{e,N}^{2}}\bm I_2.
\label{eq:single_path_fim_partition}
\end{equation}
The equivalent Fisher information for delay, Doppler, azimuth, and elevation is
\begin{equation}
\boxed{
\bm F_{\mathrm s}^{\mathrm e}
=
\bm F_{\mathrm{ss}}
-
\bm F_{\mathrm{s}\gamma}
\bm F_{\gamma\gamma}^{-1}
\bm F_{\gamma\mathrm{s}}}
\label{eq:single_path_gain_schur}
\end{equation}
and $\mathrm{Cov}(\widehat{\bm\theta}_{\mathrm s})\succeq(\bm F_{\mathrm s}^{\mathrm e})^{-1}$. For the equivalent homogeneous-noise model, active-set centering and the centered URA yield
\begin{equation}
\bm F_{\mathrm{s}\gamma}=\bm 0,
\qquad
\bm F_{\mathrm s}^{\mathrm e}=\bm F_{\mathrm{ss}},
\label{eq:single_path_gain_orthogonality}
\end{equation}
because $\sum_{\mathcal A}\widetilde f_{k,m}=\sum_{\mathcal A}\widetilde t_{k,m}=0$ and $\bm b^H\bm g_{\varphi}=\bm b^H\bm g_{\vartheta}=0$. For the exact heteroscedastic QAM likelihood, the same Schur complement applies with symbol-dependent weights and the corresponding weighted centering.

\begin{figure*}[hb]    
\begin{align}
\left\|\bm g_\varphi\right\|^2
&=
\left(2\pi\frac{\cos\vartheta}{\lambda}\right)^{\!2}
\Bigg(
d_{R,x}^2\sin^2\!\varphi\; N_{R,y}\!\sum_{m=0}^{N_{R,x}-1}\!\tilde{m}^2
+
d_{R,y}^2\cos^2\!\varphi\; N_{R,x}\!\sum_{n=0}^{N_{R,y}-1}\!\tilde{n}^2
\Bigg)
\label{eq:norm_gphi}\\
\left\|\bm g_\vartheta\right\|^2
&=
\left(2\pi\frac{\sin\vartheta}{\lambda}\right)^{\!2}
\Bigg(
d_{R,x}^2\cos^2\!\varphi\; N_{R,y}\!\sum_{m=0}^{N_{R,x}-1}\!\tilde{m}^2
+
d_{R,y}^2\sin^2\!\varphi\; N_{R,x}\!\sum_{n=0}^{N_{R,y}-1}\!\tilde{n}^2
\Bigg)
\label{eq:norm_gtheta}
\end{align}
\end{figure*}

\paragraph{Delay/Doppler FIM entries}
Define the exact active-set moments
\begin{equation}
S_{ff}\triangleq\sum_{(k,m)\in\mathcal A}\widetilde f_{k,m}^{2},\quad
S_{tt}\triangleq\sum_{(k,m)\in\mathcal A}\widetilde t_{k,m}^{2},
\label{eq:fim_summ}
\end{equation}
\begin{equation}
S_{ft}\triangleq\sum_{(k,m)\in\mathcal A}\widetilde f_{k,m}\widetilde t_{k,m}.
\label{eq:fim_summ2}
\end{equation}
The nuisance-eliminated delay/Doppler entries are
\begin{align}
F_{\tau\tau}
&=2\,\mathrm{SNR}_{\mathrm{eff}}N_R(2\pi)^2S_{ff},
\label{eq:fim_tautau}\\
F_{f_Df_D}
&=2\,\mathrm{SNR}_{\mathrm{eff}}N_R(2\pi)^2S_{tt},
\label{eq:fim_fdfd}\\
F_{\tau f_D}
&=-2\,\mathrm{SNR}_{\mathrm{eff}}N_R(2\pi)^2S_{ft}.
\label{eq:fim_taufd}
\end{align}
Thus, the delay--Doppler block is evaluated from the actual paired frequency--time coordinates and is not forced to be diagonal. For a rectangular allocation with $K$ contiguous tones repeated over $M$ equally spaced symbols,
\begin{equation}
S_{ff}=M(\Delta f)^2\frac{K(K^2-1)}{12},
\quad
S_{tt}=K T_s^2\frac{M(M^2-1)}{12},
\label{eq:rectangular_active_moments}
\end{equation}
$S_{ft}=0$. For a fixed-offset regular PRS comb with $K_{\mathrm{PRS}}$ tones per symbol, the frequency moment becomes:
\begin{equation}
S_{ff}
=M(K_{\mathrm{comb}}\Delta f)^2
\frac{K_{\mathrm{PRS}}(K_{\mathrm{PRS}}^2-1)}{12}.
\label{eq:prs_comb_frequency_moment}
\end{equation}
For the NR mapping in \eqref{eq:prs_comb_freq}, however, the symbol-dependent offsets $\kappa_m$ are retained and $S_{ff}$, $S_{tt}$, and $S_{ft}$ are computed directly from $\mathcal A_{\mathrm{PRS}}$.

\paragraph{Angle FIM entries}
Let $\bm g_\varphi\triangleq\partial\bm b/\partial\varphi$ and $\bm g_\vartheta\triangleq\partial\bm b/\partial\vartheta$. Since there are $N_{\mathcal A}$ active observations, the angular block is:
\begin{align}
F_{\varphi\varphi}
&=2\,\mathrm{SNR}_{\mathrm{eff}}N_{\mathcal A}\|\bm g_\varphi\|^2,
\label{eq:fim_phiphi}\\
F_{\vartheta\vartheta}
&=2\,\mathrm{SNR}_{\mathrm{eff}}N_{\mathcal A}\|\bm g_\vartheta\|^2,
\label{eq:fim_thetatheta}\\
F_{\varphi\vartheta}
&=2\,\mathrm{SNR}_{\mathrm{eff}}N_{\mathcal A}
\Re\!\left\{\bm g_\varphi^H\bm g_\vartheta\right\}.
\label{eq:fim_phitheta}
\end{align}
The centered URA and active-set centering also make the cross-information between $\{\tau,f_D\}$ and $\{\varphi,\vartheta\}$ vanish. Hence, $\bm F_{\mathrm s}^{\mathrm e}$ separates into a generally non-diagonal $2\times2$ delay--Doppler block and a $2\times2$ angle block.

\paragraph{Closed forms of $\|\bm g_\varphi\|^2$ and $\|\bm g_\vartheta\|^2$}
For a receive URA of size $N_{R,x}\times N_{R,y}$, the derivative norms are given in \eqref{eq:norm_gphi} and \eqref{eq:norm_gtheta}. The antenna-index sums satisfy $\sum_{m=0}^{N_{R,x}-1}\widetilde m^2=N_{R,x}(N_{R,x}^2-1)/12$ and $\sum_{n=0}^{N_{R,y}-1}\widetilde n^2=N_{R,y}(N_{R,y}^2-1)/12$.

\subsubsection{Open form CRLBs for delay, Doppler, and angles}

\begin{equation}
D_{ft}\triangleq S_{ff}S_{tt}-S_{ft}^{2}
\label{eq:delay_doppler_determinant}
\end{equation}
When $D_{ft}>0$, inversion of the exact delay--Doppler block gives
\begin{align}
\mathrm{CRLB}(\tau)
&=
\frac{S_{tt}}
{2\,\mathrm{SNR}_{\mathrm{eff}}N_R(2\pi)^2D_{ft}},
\label{eq:crlb_tau}\\
\mathrm{CRLB}(f_D)
&=
\frac{S_{ff}}
{2\,\mathrm{SNR}_{\mathrm{eff}}N_R(2\pi)^2D_{ft}}.
\label{eq:crlb_fd}
\end{align}
If Doppler is known or compensated during delay estimation, the conditional delay bound reduces to
\begin{equation}
\mathrm{CRLB}(\tau\mid f_D)
=
\frac{1}{2\,\mathrm{SNR}_{\mathrm{eff}}N_R(2\pi)^2S_{ff}}.
\label{eq:crlb_tau_doppler_known}
\end{equation}
Substitution of \eqref{eq:rectangular_active_moments} recovers the usual contiguous-tone formulas, whereas substitution of \eqref{eq:prs_comb_frequency_moment} gives the regular-comb formula with spacing $K_{\mathrm{comb}}\Delta f$. For the angles:
\begin{equation}
\Delta_{\mathrm{ang}}
\triangleq
F_{\varphi\varphi}F_{\vartheta\vartheta}-F_{\varphi\vartheta}^{2}.
\label{eq:det_ang}
\end{equation}
Then the CRLB can represented as:
\begin{align}
\mathrm{CRLB}(\varphi)
&=
\frac{\|\bm g_{\vartheta}\|^{2}}
{2N_{\mathcal A}\,\mathrm{SNR}_{\mathrm{eff}}
\left(\|\bm g_{\varphi}\|^{2}\|\bm g_{\vartheta}\|^{2}
-\Re\!\left\{\bm g_{\varphi}^{H}\bm g_{\vartheta}\right\}^{2}\right)}
\label{eq:crlb_phi}\\
\mathrm{CRLB}(\vartheta)
&=
\frac{\|\bm g_{\varphi}\|^{2}}
{2N_{\mathcal A}\,\mathrm{SNR}_{\mathrm{eff}}
\left(\|\bm g_{\varphi}\|^{2}\|\bm g_{\vartheta}\|^{2}
-\Re\!\left\{\bm g_{\varphi}^{H}\bm g_{\vartheta}\right\}^{2}\right)}
\label{eq:crlb_theta}
\end{align}
If $F_{\varphi\vartheta}$ is negligible, these simplify to $1/(2N_{\mathcal A}\mathrm{SNR}_{\mathrm{eff}}\|\bm g_\varphi\|^2)$ and $1/(2N_{\mathcal A}\mathrm{SNR}_{\mathrm{eff}}\|\bm g_\vartheta\|^2)$.

\paragraph{Range/velocity scaling (optional)}
For monostatic ranging $R=\frac{c\tau}{2}$ and Doppler-to-velocity relation $v=\frac{c f_D}{2f_c}$, the variance-scaling identities are
\begin{equation}
\mathrm{CRLB}(R)=\left(\frac{c}{2}\right)^2\mathrm{CRLB}(\tau),
\label{eq:range_vel_scaling}
\end{equation}

\begin{equation}
\mathrm{CRLB}(v)=\left(\frac{c}{2f_c}\right)^2\mathrm{CRLB}(f_D)
\label{eq:range_vel_scaling2}
\end{equation}

\begin{figure*}[hb]
    \centering
    \includegraphics[width=1\linewidth, height=0.35\linewidth]{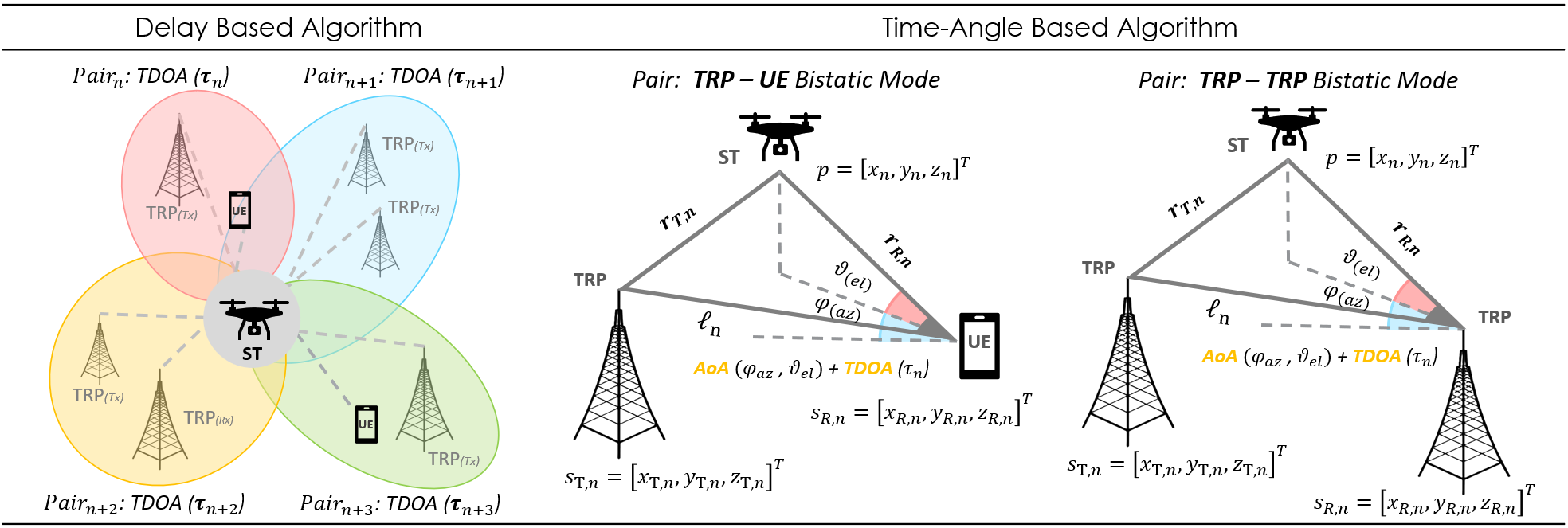}
    \caption{Illustration of two positioning methods using different bistatic sensing pairs. Each pair contains the illumination link $\mathrm{Tx}\!\rightarrow\!\mathrm{ST}$, the echo link $\mathrm{ST}\!\rightarrow\!\mathrm{Rx}$, and the direct baseline link $\mathrm{Tx}\!\rightarrow\!\mathrm{Rx}$. A TRP or UE receiver defines the TRP--TRP or TRP--UE mode, respectively. Rx estimates the LOS-referenced excess delay $\Delta\tau_n$ and target-echo azimuth/elevation; the ``TDOA($\tau_n$)'' labels in the schematic are shorthand for $\Delta\tau_n$ in \eqref{eq:tdoa_direct_path}.}
    \label{Time_Angle}
\end{figure*}

\section{Positioning Algorithms and Error Bounds}
\label{sec:postioning}
\

\

The simulation framework employs two bistatic positioning algorithms. Their corresponding \emph{3D} positioning error bounds (PEBs) are obtained from the CRLBs of the intermediate sensing parameters in Section~\ref{sec:boundary_analysis}. The target position belongs to $\mathbb{R}^{3}$.

\subsection{Positioning Algorithms}
\label{subsec:positioning_algorithms}

Consider a set $\mathcal{V}$ of selected bistatic links. Link $n\in\mathcal{V}$ is formed by a Tx at position $\bm s_{T,n}=[x_{T,n},y_{T,n},z_{T,n}]^T$ and a Rx at position $\bm s_{R,n}=[x_{R,n},y_{R,n},z_{R,n}]^T$, while the passive sensing target is at the unknown position $\bm p=[x,y,z]^T$. 
Define the Tx--target, Rx--target, and Tx--Rx distances as:
\begin{equation}
r_{T,n} \triangleq \|\bm p-\bm s_{T,n}\|,\
r_{R,n} \triangleq \|\bm p-\bm s_{R,n}\|,\
\ell_n \triangleq \|\bm s_{T,n}-\bm s_{R,n}\|
\label{eq:dist_def}
\end{equation}
The bistatic delay of the target echo observed on link $n$ is:
\begin{equation}
\tau_n = \frac{r_{T,n}+r_{R,n}}{c}
\label{eq:bistatic_delay}
\end{equation}
where $c$ is the speed of light. If the direct Tx$\rightarrow$Rx path is also observed, its delay is $\tau_{\mathrm{LOS},n}=\ell_n/c$, and a synchronization-robust excess delay (or TDOA with respect to the LOS path) can be formed as:
\begin{equation}
\Delta\tau_n \triangleq \tau_n-\tau_{\mathrm{LOS},n}
=
\frac{r_{T,n}+r_{R,n}-\ell_n}{c}
\label{eq:tdoa_direct_path}
\end{equation}
Equivalently, the corresponding bistatic range sum is:
\begin{equation}
\bar r_n \triangleq r_{T,n}+r_{R,n}=c\tau_n=\ell_n+c\,\Delta\tau_n
\label{eq:bistatic_range_sum}
\end{equation}

\vspace{-12pt}

\paragraph{Intermediate parameter estimation (delay and DoA)}
Under the normalized OFDM array observation model in \eqref{eq:zkm_normalized}, each receiver first estimates the intermediate parameters (excess delay and 3D DoA) and then converts these estimates into a Cartesian position using the geometry in \eqref{eq:dist_def}--\eqref{eq:pos_est_time_angle}. Let $\mathcal A_n$ denote the active resource-element set available on pair $n$. For DoA estimation, the sample spatial covariance matrix is formed only from those active observations:
\begin{equation}
\bm R_{z,n}
\triangleq
\frac{1}{N_{\mathcal A,n}}
\sum_{(k,m)\in\mathcal A_n}
\bm z_n[k,m]\bm z_n^{H}[k,m],
\quad
N_{\mathcal A,n}=|\mathcal A_n|
\label{eq:cov_doa}
\end{equation}
An eigendecomposition of $\bm R_{z,n}$ is then performed, and the 2D MUSIC spectrum over azimuth and elevation is evaluated:
\begin{equation}
P_{\mathrm{MUSIC}}(\varphi,\vartheta)
=
\frac{1}{\bm b^{H}(\varphi,\vartheta)\,\bm U_{n}\bm U_{n}^{H}\,\bm b(\varphi,\vartheta)}
\label{eq:music_2d}
\end{equation}
where $\bm U_n$ spans the noise subspace. The DoA estimate is obtained from the peak of \eqref{eq:music_2d}. For delay estimation, after receive-side spatial combining using the estimated DoA,
$\tilde z_n[k,m]\triangleq \bm b^{H}(\widehat\varphi_n,\widehat\vartheta_n)\bm z_n[k,m]$, the delay is estimated over the same active set as
\begin{equation}
\widehat\tau_n
=
\arg\max_{\tau}
\left|
\sum_{(k,m)\in\mathcal A_n}
\tilde z_n[k,m]\;e^{j2\pi f_{k,n}\tau}
\right|^{2}
\label{eq:delay_periodogram}
\end{equation}
where $f_{k,n}$ is the actual baseband frequency of active subcarrier $k$ on pair $n$. The LOS-referenced estimate used by both positioning algorithms is $\widehat{\Delta\tau}_n=\widehat\tau_n-\widehat\tau_{\mathrm{LOS},n}$, with $\widehat\tau_{\mathrm{LOS},n}$ obtained from the direct-path peak. Thus, a PRS comb is processed at its mapped frequencies rather than at fictitious contiguous indices.

\vspace{-18pt}
\subsubsection{Time-Angle Based Algorithm (Excess Delay + DoA)}
\label{subsubsec:alg_time_angle}
\vspace{4pt}
The algorithm first forms a closed-form position estimate for each valid pair and then jointly refines one common target position using all selected range-sum and DoA measurements. The per-pair estimates are used only to construct a stable initializer and are not averaged to form the final output.
\vspace{-8pt}
\paragraph{Step 1: 3D arrival direction.}
From the estimated azimuth--elevation pair $(\varphi_n,\vartheta_n)$, the unit vector pointing from receiver $n$ towards the target:
\begin{equation}
\bm u_n(\varphi_n,\vartheta_n)
\triangleq
\begin{bmatrix}
\cos\vartheta_n\cos\varphi_n\\
\cos\vartheta_n\sin\varphi_n\\
\sin\vartheta_n
\end{bmatrix}
\label{eq:u_target}
\end{equation}
The Tx--Rx baseline direction, expressed at the receiver:
\begin{equation}
\bm u_{\mathrm{LOS},n}
\triangleq
\frac{\bm s_{T,n}-\bm s_{R,n}}{\ell_n}
\label{eq:u_los}
\end{equation}
In the implementation this direction can either be computed from the known node coordinates or recovered from the LOS peak. Hence, the included angle between the direct-path direction and the target-echo direction is $\cos\psi_n
\triangleq
\bm u_{\mathrm{LOS},n}^{T}\bm u_n$.

\paragraph{Step 2: Closed-form range from the receiver to the target.}
Using the triangle formed by the transmitter, receiver, and target together with $\bar r_n=r_{T,n}+r_{R,n}$, the receiver--target distance is obtained from the law of cosines as:
\begin{equation}
r_{R,n}
=
\frac{\bar r_n^{2}-\ell_n^{2}}
{2\left(\bar r_n-\ell_n\cos\psi_n\right)}
\label{eq:range_from_time_angle}
\end{equation}
The Tx--ST distance then follows from $r_{T,n}=\bar r_n-r_{R,n}$.

\paragraph{Step 3: 3D position estimate and joint multi-link fusion.}
The closed-form position estimate obtained from link $n$:
\begin{equation}
\widehat{\bm p}_n
=
\bm s_{R,n} + \widehat r_{R,n}\,\bm u_n(\widehat\varphi_n,\widehat\vartheta_n),
\label{eq:pos_est_time_angle}
\end{equation}
where $\widehat r_{R,n}$ is computed by inserting $(\widehat{\bar r}_n,\widehat\varphi_n,\widehat\vartheta_n)$ into \eqref{eq:range_from_time_angle}--\eqref{eq:u_target}. For multiple valid links, their arithmetic mean is used only as the common nonlinear-solver initializer,
\begin{equation}
\widehat{\bm p}_{\mathrm{TA}}^{(0)}
=
\frac{1}{|\mathcal{V}_{\mathrm{TA}}|}
\sum_{n\in\mathcal{V}_{\mathrm{TA}}}
\widehat{\bm p}_n.
\label{eq:ta_initializer}
\end{equation}
Define the bistatic range-sum residual as
\begin{equation}
e_{r,n}(\bm p)
\triangleq
\|\bm p-\bm s_{T,n}\|+\|\bm p-\bm s_{R,n}\|-\widehat{\bar r}_n.
\label{eq:ta_range_residual}
\end{equation}
The final Time--Angle fusion is implemented in two stages. Let $\widehat{\bm u}_n=\bm u_n(\widehat\varphi_n,\widehat\vartheta_n)$ denote the rx-side target direction and define the orthogonal DoA-ray projector
\begin{equation}
\bm P_{\perp,n}
\triangleq
\bm I_3-\widehat{\bm u}_n\widehat{\bm u}_n^{T}.
\label{eq:ta_doa_projector}
\end{equation}
The first stage obtains a stable common geometric estimate by fitting all selected range-sum and DoA-ray constraints jointly:
\begin{align}
J_{\mathrm g}(\bm p)
&\triangleq
\frac{1}{4|\mathcal{V}_{\mathrm{TA}}|}
\sum_{n\in\mathcal{V}_{\mathrm{TA}}}
\left[
 e_{r,n}^{2}(\bm p)
 +\frac{1}{2}
 \left\|
 \bm P_{\perp,n}(\bm p-\bm s_{R,n})
 \right\|_{2}^{2}
\right],
\nonumber\\
\widehat{\bm p}_{\mathrm g}
&=
\arg\min_{\bm p\in\mathcal B}J_{\mathrm g}(\bm p).
\label{eq:ta_geometric_objective}
\end{align}
Both residual blocks entering $J_{\mathrm g}$ are expressed in meters. The DoA-ray term avoids the azimuth singularity near the vertical direction and, unlike the former arithmetic average, jointly enforces consistency among all links. Equation~\eqref{eq:ta_geometric_objective} is solved from the common seed in \eqref{eq:ta_initializer}, a delay-ellipsoid least-squares seed when at least four valid range-sum measurements are available, and the participating-node centroid. The delay-only objective uses the same multi-start candidates; its optional Time--Angle seed affects initialization only and never enters the delay residuals or weights.

All covariance quantities used in the second stage are evaluated at the single common estimate $\widehat{\bm p}_{\mathrm g}$, not at separate closed-form pair estimates. The plug-in excess-delay variance $\widehat\sigma_{\Delta\tau,n}^{2}$ is obtained from the same joint LOS--target delay covariance used in \eqref{eq:tdoa_var}. If the local two-path FIM is ill-conditioned, the implementation uses a conservatively inflated sum of the unequal-power isolated LOS and target delay bounds. The angular covariance $\widehat{\bm C}_{\Omega,n}$ is obtained from the nuisance-eliminated angular information associated with \eqref{eq:angle_var_mode_general}. The configured reference SNR, physical link-budget model, exact active resource-element set, and the corresponding $2\!\times\!2$ UE or $8\!\times\!8$ TRP receive array are used in both calculations; the true target coordinates are never supplied to the weighting stage. Delay and MUSIC grid resolution floors are then added. In the Monte-Carlo implementation, the angular matrix is represented by its trace-equivalent isotropic covariance before link-to-link variance-ratio regularization. This retains SNR and aperture-dependent reliability difference while preventing a locally overconfident covariance eigen-direction from dominating the solution. Thus,
\begin{equation}
\widetilde\sigma_{r,n}^{2}
=
c^{2}\widehat\sigma_{\Delta\tau,n}^{2}
+\sigma_{r,\mathrm{grid}}^{2},
\qquad
\overline{\bm C}_{\Omega,n}
=
\frac{1}{2}\operatorname{tr}\!\left(\widehat{\bm C}_{\Omega,n}\right)\bm I_2.
\label{eq:ta_regularized_covariances}
\end{equation}

Let $\bm E_n=[\bm e_{\varphi,n},\bm e_{\vartheta,n}]\in\mathbb R^{3\times2}$ be an orthonormal tangent basis satisfying $\bm E_n^{T}\widehat{\bm u}_n=\bm 0$ and $\bm E_n^{T}\bm E_n=\bm I_2$. The two-dimensional lateral DoA-ray residual is
\begin{equation}
\bm q_n(\bm p)
\triangleq
\bm E_n^{T}(\bm p-\bm s_{R,n}).
\label{eq:ta_angle_residuals}
\end{equation}
With the nominal range $\widehat r_{\mathrm g,n}=\|\widehat{\bm p}_{\mathrm g}-\bm s_{R,n}\|$ clipped to the receiver-to-point distance interval induced by $\mathcal B$ and $\bm D_n=\diag\!\left(\max\{\cos\widehat\vartheta_n,c_{\min}\},1\right)$, where $c_{\min}>0$ is a numerical floor, its lateral covariance is approximated:
\begin{equation}
\widetilde{\bm C}_{q,n}
=
\widehat r_{\mathrm g,n}^{2}
\bm D_n\overline{\bm C}_{\Omega,n}\bm D_n
+\sigma_{q,\min}^{2}\bm I_2,
\quad
\bm L_{q,n}\bm L_{q,n}^{T}=\widetilde{\bm C}_{q,n}.
\label{eq:ta_lateral_covariance}
\end{equation}
The robust covariance-weighted candidate is:
\begin{equation}
\widehat{\bm p}_{\mathrm W}
=
\arg\min_{\bm p\in\mathcal B}
\sum_{n\in\mathcal{V}_{}}
\rho_{\kappa}\!\left[
\frac{|e_{r,n}(\bm p)|}{\sqrt{\widetilde\sigma_{r,n}^{2}}}
+
\left\|\bm L_{q,n}^{-1}\bm q_n(\bm p)\right\|_2
\right]
\label{eq:ta_joint_objective}
\end{equation}
where $\rho_{\kappa}(a)=\tfrac{1}{2}a^{2}$ for $a\leq\kappa$ and $\rho_{\kappa}(a)=\kappa a-\tfrac{1}{2}\kappa^{2}$ otherwise. Separate block-Huber losses are applied to the normalized range residual and to the norm of the whitened two-dimensional DoA residual, in agreement with the implementation. The objective is solved from $\widehat{\bm p}_{\mathrm g}$ using scaled damped Gauss--Newton/Levenberg--Marquardt iterations, bounded steps, and projection onto $\mathcal B$. A non-oracle geometric safeguard prevents the covariance-weighted stage from replacing the stable joint solution with a spurious local minimum:
\begin{equation}
\widehat{\bm p}_{\mathrm{TA}}
=
\begin{cases}
\widehat{\bm p}_{\mathrm W},\qquad

J_{\mathrm g}(\widehat{\bm p}_{\mathrm W})
\leq
\gamma_{\mathrm g}J_{\mathrm g}(\widehat{\bm p}_{\mathrm g})+\epsilon_{\mathrm g} \\
\;\land\; 
\|\widehat{\bm p}_{\mathrm W}-\widehat{\bm p}_{\mathrm g}\|\leq d_{\max},
\\
\widehat{\bm p}_{\mathrm g}, \qquad  \text{otherwise}.
\end{cases}
\label{eq:ta_geometric_guard}
\end{equation}

\begin{algorithm}[t]
\caption{Robust variance-weighted joint Time--Angle positioning from bistatic pairs}
\label{alg:time_angle}
\begin{algorithmic}[1]
\STATE \textbf{Input:} valid pair set $\mathcal{V}_{\mathrm{TA}}$, node coordinates $\{\bm s_{T,n},\bm s_{R,n}\}$, active observations, receiver-array sizes, reference-SNR/link-budget parameters, and fact. $\mathcal B$
\STATE \textbf{Output:} joint 3D estimate $\widehat{\bm p}_{\mathrm{TA}}$
\STATE Initialize the closed-form seed set $\mathcal P\leftarrow\emptyset$ and valid measurement set $\mathcal M\leftarrow\emptyset$
\FOR{each pair $n\in\mathcal{V}_{\mathrm{TA}}$}
    \STATE Estimate $(\widehat\varphi_n,\widehat\vartheta_n)$ from \eqref{eq:music_2d} and $\widehat{\Delta\tau}_n$ from the LOS-cancelled delay profile
    \STATE Set $\widehat{\bar r}_n=\ell_n+c\widehat{\Delta\tau}_n$ and compute the closed-form pair seed $\widehat{\bm p}_n$ from \eqref{eq:range_from_time_angle}--\eqref{eq:pos_est_time_angle}
    \STATE Append valid $\widehat{\bm p}_n$ to $\mathcal P$ and $(\widehat{\bar r}_n,\widehat{\bm u}_n)$, the active set, and the receive-array size to $\mathcal M$
\ENDFOR
\IF{$|\mathcal M|=0$}
    \STATE \textbf{return} positioning failure
\ENDIF
\STATE Form $\widehat{\bm p}_{\mathrm{TA}}^{(0)}$ and the common multi-start set from it, the delay-ellipsoid seed when available, and the participating-node centroid
\STATE Minimize $J_{\mathrm g}$ in \eqref{eq:ta_geometric_objective} from every valid seed and retain the lowest-cost joint geometric solution $\widehat{\bm p}_{\mathrm g}$
\STATE At the same $\widehat{\bm p}_{\mathrm g}$ for every link, evaluate $\widetilde\sigma_{r,n}^{2}$ and $\overline{\bm C}_{\Omega,n}$; apply the conservative delay fallback, resolution floors, and bounded inter-link variance ratios
\STATE Form $\bm E_n$, $\bm q_n(\bm p)$, and $\widetilde{\bm C}_{q,n}$ using \eqref{eq:ta_angle_residuals}--\eqref{eq:ta_lateral_covariance}
\STATE Starting from $\widehat{\bm p}_{\mathrm g}$, min. block-Huber objective in \eqref{eq:ta_joint_objective} over $\mathcal B$ using scaled damped LM iterations to $\widehat{\bm p}_{\mathrm W}$
\STATE Apply the non-oracle safeguard in \eqref{eq:ta_geometric_guard}
\STATE \textbf{return} $\widehat{\bm p}_{\mathrm{TA}}$
\end{algorithmic}
\end{algorithm}

\begin{algorithm}[t]
\caption{Delay Based - 3D ellipsoid multilateration}
\label{alg:delay_only}
\begin{algorithmic}[1]
\STATE \textbf{Input:} valid pair set $\mathcal{V}_{\mathrm{D}}$, coordinates $\{\bm s_{T,n},\bm s_{R,n}\}$, delay measurements $\{\widehat{\Delta\tau}_n\}$, optional weights $\{w_n\}$
\STATE \textbf{Output:} 3D estimate $\widehat{\bm p}_{\mathrm{D}}$
\IF{$|\mathcal{V}_{\mathrm{D}}|<4$}
    \STATE \textbf{return} positioning failure
\ENDIF
\FOR{each pair $n\in\mathcal{V}_{\mathrm{D}}$}
    \STATE Form the bistatic range sum $\widehat{\bar r}_n=\ell_n+c\,\widehat{\Delta\tau}_n$
\ENDFOR
\STATE Form $\mathcal P_0$ from the optional Time--Angle seed (initialization only), the delay-ellipsoid least-squares seed, and the participating-node centroid
\STATE Initialize the best cost $J_{\mathrm{best}}\leftarrow\infty$
\FOR{each valid seed $\bm p^{(0)}\in\mathcal P_0$}
    \STATE Set the iteration index $i\leftarrow 0$
    \REPEAT
        \STATE Evaluate the residual vector $\bm e(\bm p^{(i)})$ using \eqref{eq:residual_def}
        \STATE Form the Jacobian rows $\bm j_n^{T}(\bm p^{(i)})$ using \eqref{eq:jacobian_tdoa}
        \STATE Update the position through the damped Gauss--Newton / Levenberg--Marquardt step in \eqref{eq:gauss_newton}
        \STATE Set $i\leftarrow i+1$
    \UNTIL{The stopping condition or maximum number of iterations is reached}
    \STATE Retain $\bm p^{(i)}$ if its delay-residual cost < $J_{\mathrm{best}}$
\ENDFOR
\STATE \textbf{return} the lowest-cost estimate $\widehat{\bm p}_{\mathrm{D}}$
\end{algorithmic}
\end{algorithm}

The safeguard depends only on measured range/DoA consistency and the known factory volume. The numerical implementation uses a resolved-path fallback inflation factor of $2$, a block-Huber threshold $\kappa=2.5$, a maximum inter-link variance ratio of $25$, a $0.5^{\circ}$ angular standard-deviation floor, and a $0.25$~m lateral floor. The safeguard parameters are $\gamma_{\mathrm g}=1.15$, $\epsilon_{\mathrm g}=0.25$~m$^2$, and $d_{\max}=50$~m. Hence, the arithmetic average in \eqref{eq:ta_initializer} remains only one initializer, while both the Time--Angle and delay-only pipelines perform joint multi-link nonlinear fusion. Algorithm~\ref{alg:time_angle} summarizes the implementation.

\vspace{-1.5cm}
\subsubsection{Delay-Based Algorithm}
\label{subsubsec:alg_tdoa}

The algorithm relies only on delay measurements. Each measured bistatic range sum defines a 3D ellipsoid whose foci are the Tx and Rx of the corresponding bistatic link.
\paragraph{Ellipsoid measurement model.}
For every selected link $n\in\mathcal{V}_{\mathrm{D}}$, the measured excess delay yields the bistatic range sum $\widehat{\bar r}_n=\ell_n+c\,\widehat{\Delta\tau}_n$, which imposes the constraint
\begin{equation}
\|\bm p-\bm s_{T,n}\|+\|\bm p-\bm s_{R,n}\|
=
\widehat{\bar r}_n,
\qquad n\in\mathcal{V}_{\mathrm{D}}.
\label{eq:ellipsoid_constraint}
\end{equation}
Hence, the target must lie near the intersection of several ellipsoids in $\mathbb{R}^{3}$. In general, at least four non-coplanar links are required for unique 3D positioning.

\paragraph{Least-squares formulation.}
Rather than intersecting only four ellipsoids exactly, the implementation exploits all valid links and estimates the target as the minimizer of the residual mismatch
\begin{equation}
\widehat{\bm p}_{_\mathrm{D}}
=
\arg\min_{\bm p\in\mathbb{R}^{3}}
\sum_{n}
 w_n
\Big[\|\bm p-\bm s_{T,n}\|+\|\bm p-\bm s_{R,n}\|-\widehat{\bar r}_n
\Big]^{2}
\label{eq:ellipsoid_ls}
\end{equation}
where $w_n\ge 0$ are optional weights. Uniform weights are used in the base implementation, while $w_n=1/\sigma_{\Delta\tau,n}^{2}$ provides the weighted-LS counterpart when per-link delay variances are available.

\paragraph{Gauss--Newton / Levenberg--Marquardt refinement.}
Define the residual of link $n$ as:
\begin{equation}
e_n(\bm p)
\triangleq
\|\bm p-\bm s_{T,n}\|+\|\bm p-\bm s_{R,n}\|-\widehat{\bar r}_n,
\qquad n\in\mathcal{V}_{\mathrm{D}},
\label{eq:residual_def}
\end{equation}
and stack terms in $\bm e(\bm p)$. The Gauss--Newton update:
\begin{equation}
\bm p^{(i+1)}
=
\bm p^{(i)}-
\left(\bm J^T\bm W\bm J+\lambda^{(i)}\bm I_3\right)^{-1}
\bm J^T\bm W\,\bm e(\bm p^{(i)}),
\label{eq:gauss_newton}
\end{equation}
where $\bm W=\diag(w_n)$ and $\lambda^{(i)}\ge 0$ is the damping parameter. The $n$th row of the Jacobian is
\begin{equation}
\bm j_n^T(\bm p)
=
\left(
\frac{\bm p-\bm s_{T,n}}{\|\bm p-\bm s_{T,n}\|}
+
\frac{\bm p-\bm s_{R,n}}{\|\bm p-\bm s_{R,n}\|}
\right)^T
\label{eq:jacobian_tdoa}
\end{equation}
In practice, the algorithm uses the same multi-start candidates as the joint Time--Angle refinement: $\widehat{\bm p}_{\mathrm{TA}}^{(0)}$, the delay-ellipsoid least-squares seed when available, and the participating-node centroid. The Time--Angle estimate is only an optional numerical initializer; candidate selection and refinement use the delay-residual cost exclusively. Thus, ``delay-only'' refers to the measurement objective, not to the origin of every starting point. The valid solution with the smallest delay-residual cost is retained, as summarized in Algorithm~\ref{alg:delay_only}.

\begin{figure*}[ht]
\begin{equation}
\|\bm g_{\varphi,n}\|^2
=
\left(2\pi\frac{\cos\vartheta_n}{\lambda}\right)^{\!2}
\frac{N_{R,x,n}N_{R,y,n}}{12}
\left[
 d_{R,x,n}^{2}(N_{R,x,n}^{2}-1)\sin^{2}\varphi_n
 +
 d_{R,y,n}^{2}(N_{R,y,n}^{2}-1)\cos^{2}\varphi_n
\right]
\label{eq:crn_array_const}
\end{equation}
\begin{equation}
\|\bm g_{\vartheta,n}\|^2
=
\left(2\pi\frac{\sin\vartheta_n}{\lambda}\right)^{\!2}
\frac{N_{R,x,n}N_{R,y,n}}{12}
\left[
 d_{R,x,n}^{2}(N_{R,x,n}^{2}-1)\cos^{2}\varphi_n
 +
 d_{R,y,n}^{2}(N_{R,y,n}^{2}-1)\sin^{2}\varphi_n
\right]
\label{eq:g_norm_square_general}
\end{equation}
\end{figure*}

\begin{figure*}[hb]
\begin{equation}
\begin{aligned}
\bm I_{\mathrm{TA},m}(\bm p)
=
\sum_{n\in\mathcal{V}_{m}^{\mathrm{TA}}}
\Bigg(
\frac{1}{c^2\sigma_{\Delta\tau,n}^{2}}\,
\bm j_n(\bm p)\bm j_n^{T}(\bm p)
+\frac{1}{\sigma_{\varphi,n}^{2}}\,
\frac{\partial\varphi_n}{\partial\bm p}
\left(\frac{\partial\varphi_n}{\partial\bm p}\right)^{T}
+
\frac{1}{\sigma_{\vartheta,n}^{2}}\,
\frac{\partial\vartheta_n}{\partial\bm p}
\left(\frac{\partial\vartheta_n}{\partial\bm p}\right)^{T}
\Bigg)
\end{aligned}
\label{eq:fim_ta_mode_open}
\end{equation}
\end{figure*}

\begin{figure*}[hb]

\begin{equation}\boxed{
\mathrm{PEB}(\bm I)
=
\sqrt{
\frac{
I_{yy}I_{zz}-I_{yz}^{2}
+
I_{xx}I_{zz}-I_{xz}^{2}
+
I_{xx}I_{yy}-I_{xy}^{2}
}{
I_{xx}I_{yy}I_{zz}
+
2I_{xy}I_{xz}I_{yz}
-
I_{xx}I_{yz}^{2}
-
I_{yy}I_{xz}^{2}
-
I_{zz}I_{xy}^{2}
}
}}
\label{eq:peb_open_3d}
\end{equation}
\end{figure*}

\subsection{Positioning Error Bounds of Bistatic Modes}
\label{subsec:peb_bistatic}

A bistatic sensing pair $n$ is the ordered tuple $(\mathrm{Tx}_n,\mathrm{ST},\mathrm{Rx}_n)$ and is composed of the three links $\mathrm{Tx}_n\!\rightarrow\!\mathrm{ST}$, $\mathrm{ST}\!\rightarrow\!\mathrm{Rx}_n$, and the direct baseline link $\mathrm{Tx}_n\!\rightarrow\!\mathrm{Rx}_n$.
Accordingly, a pair belongs to the TRP--TRP bistatic mode when $\mathrm{Rx}_n$ is a TRP, and it belongs to the TRP--UE bistatic mode when $\mathrm{Rx}_n$ is a UE.
Hybrid sensing combines pairs from both modes.
Let $\mathcal{V}_{\mathrm{TT}}$, $\mathcal{V}_{\mathrm{TU}}$, $\mathcal{V}_{\mathrm{H}}\triangleq \mathcal{V}_{\mathrm{TT}}\cup\mathcal{V}_{\mathrm{TU}}$ denote the feasible pair universes for the TRP--TRP, TRP--UE, and Hybrid sensing modes, respectively. Let $\mathcal{S}_{m}(\widehat{\bm p}_{\mathrm{sch}})\subseteq\mathcal{V}_{m}$ denote the scheduled pair set for mode $m\in\{\mathrm{TT},\mathrm{TU},\mathrm{H}\}$. The exact target position is unavailable to the scheduler. Instead, all modes receive the same coarse three-dimensional estimate $\widehat{\bm p}_{\mathrm{sch}}$, representing the output of a common initial acquisition stage or a previous tracking occasion. The simulation model is $\widehat{\bm p}_{\mathrm{sch}}=\bm p+\bm e$, where the three components of $\bm e$ are random and the estimate is projected onto the known factory volume; consequently, neither the true position nor the fixed target height is supplied to pair selection. For candidate pair $n$, the scheduling score is the estimated bistatic path length $\widehat d_{\mathrm{bi},n}=\|\widehat{\bm p}_{\mathrm{sch}}-\bm s_{T,n}\|+\|\widehat{\bm p}_{\mathrm{sch}}-\bm s_{R,n}\|$. Each baseline mode selects the $N_{\mathrm{pair}}$ pairs having the smallest scores in its own feasible universe. 

Hybrid sensing does not rank the union $\mathcal V_{\mathrm H}$ as one larger pool. Instead, the TRP--UE and TRP--TRP universes are ranked separately and the selected-link budget is partitioned by a fixed near-$50/50$ quota. For odd $N_{\mathrm{pair}}$, the $\lceil N_{\mathrm{pair}}/2\rceil$/$\lfloor N_{\mathrm{pair}}/2\rfloor$ allocation is alternated across target/Monte-Carlo realizations. Every mode therefore uses exactly $|\mathcal{S}_{m}|=N_{\mathrm{pair}}$ links and the same per-pair waveform and resource configuration. The same $\widehat{\bm p}_{\mathrm{sch}}$ is used by all modes for a given realization and is held common across waveform and SNR comparisons. In the PEB evaluation, this same practical rule determines the scheduled set, whereas the FIM itself is evaluated at the physical parameter point $\bm p$. For each sensing mode, let $\mathcal{V}_{m}^{\mathrm{TA}}\subseteq\mathcal{S}_{m}$ and $\mathcal{V}_{m}^{\mathrm{D}}\subseteq\mathcal{S}_{m}$ denote the valid-pair sets used by the time--angle and the delay algorithms, respectively. The PEB is defined as:
\begin{equation}
\mathrm{PEB}
\triangleq
\sqrt{\mathrm{tr}\!\left(\mathrm{CRLB}(\bm p)\right)}, \quad \bm p=[x,y,z]^T 
\label{eq:peb_def}
\end{equation}
For pair $n$, the equivalent effective target-echo SNR of the adopted unweighted reciprocal receiver is
\begin{equation}
\mathrm{SNR}_{\mathrm{tar,eff},n}
\triangleq
\frac{|\gamma_{\mathrm{tar},n}|^2}{\eta_n\sigma_{N,\mathrm{RE},n}^{2}}
=
\frac{P_{\mathrm{RE},n}G_{\mathrm{bi},n}|\beta_n|^2}
{\eta_n\sigma_{N,\mathrm{RE},n}^{2}}
\label{eq:snr_eff_pair}
\end{equation}
where the exact equation of the $G_{\mathrm{bi},n}$ can be derived as
$G_{\mathrm{bi},n}=G_{T,n}G_{R,n}\lambda^2\sigma_{\mathrm{RCS}}/[(4\pi)^3d_{T,n}^2d_{R,n}^2]$, and also the variance
$\sigma_{N,\mathrm{RE},n}^{2}=k_{\mathrm B}T_0F_n\Delta f_n$, and $\eta_n=1$ for PRS/BPSK/QPSK. For higher-order QAM, $\eta_n$ denotes the frame-wise or ensemble average in \eqref{eq:effective_noise2}; hence, the corresponding PEBs remain receiver-aligned average-noise benchmarks rather than the modulation-independent fundamental limits. The direct-path effective SNR is evaluated with the same per-RE transmit power and receiver noise as
\begin{equation}
\mathrm{SNR}_{\mathrm{LOS,eff},n}
\triangleq
\frac{|\gamma_{\mathrm{LOS},n}|^2}{\eta_n\sigma_{N,\mathrm{RE},n}^{2}}
=
\frac{P_{\mathrm{RE},n}G_{\mathrm{LOS},n}|\beta_{\mathrm{LOS},n}|^2}
{\eta_n\sigma_{N,\mathrm{RE},n}^{2}}.
\label{eq:snr_eff_los_pair}
\end{equation}
Moreover, $N_{R,n}=N_{R,x,n}N_{R,y,n}$ denotes the number of receive antenna elements of pair $n$, while $\mathcal A_n$ and $N_{\mathcal A,n}=|\mathcal A_n|$ denote its active resource-element set and cardinality.

For the excess-delay bound, the target and direct-path delays are not assigned identical independent errors. Both components are estimated from the same normalized OFDM-array observation. After Doppler compensation over the delay-estimation interval, the two-path mean on active resource element $(k,m)$ is written as
\begin{equation}
\begin{aligned}
\bm\mu_{n}^{(2)}[k,m]
={}&
\gamma_{\mathrm{LOS},n}
 e^{-j2\pi\widetilde f_{k,m,n}\tau_{\mathrm{LOS},n}}
 \bm b_{\mathrm{LOS},n}
\\[-1mm]
&+
\gamma_{\mathrm{tar},n}
 e^{-j2\pi\widetilde f_{k,m,n}\tau_{n}}
 \bm b_{\mathrm{tar},n},
\quad (k,m)\in\mathcal A_n
\end{aligned}
\label{eq:joint_two_path_mean}
\end{equation}
where $\widetilde f_{k,m,n}=f_{k,n}-\overline f_n$ is centered over all $(k,m)\in\mathcal A_n$, so repeated appearances of a frequency in different PRS symbols are counted explicitly; $\bm b_{\mathrm{LOS},n}$ points from the receiver toward the transmitter, and $\bm b_{\mathrm{tar},n}$ points from the receiver toward the target. The path amplitudes satisfy
\begin{equation}
\begin{aligned}
|\gamma_{\mathrm{LOS},n}|^2
&=P_{\mathrm{RE},n}G_{\mathrm{LOS},n}|\beta_{\mathrm{LOS},n}|^2,
\\
|\gamma_{\mathrm{tar},n}|^2
&=P_{\mathrm{RE},n}G_{\mathrm{bi},n}|\beta_n|^2,
\end{aligned}
\label{eq:joint_two_path_powers}
\end{equation}
so the generally much stronger LOS component and the weaker target echo enter with different powers. Define the real-valued joint parameter vector:
\begin{equation}
\begin{aligned}
\bm\xi_n
\triangleq
[\tau_{\mathrm{LOS},n},\,\tau_n,\,
\Re\{\gamma_{\mathrm{LOS},n}\},\,\Im\{\gamma_{\mathrm{LOS},n}\},\, 
&
\\
\Re\{\gamma_{\mathrm{tar},n}\},\,\Im\{\gamma_{\mathrm{tar},n}\}]^{T}
\label{eq:joint_two_path_parameter}
\end{aligned}
\end{equation}
Under the equivalent receiver-noise model in \eqref{eq:effective_noise2}, its FIM:
\begin{equation}
[\bm J_{\xi,n}]_{i,j}
=
\frac{2}{\overline{\sigma}_{e,N,n}^{2}}
\Re\!\left\{
\sum_{(k,m)\in\mathcal A_n}
\frac{\partial \bm\mu_{n}^{(2)H}{_{[k,m]}}}{\partial \xi_{n,i}}
\frac{\partial \bm\mu_{n}^{(2)}{_{[k,m]}}}{\partial \xi_{n,j}}
\right\}
\label{eq:joint_two_path_fim}
\end{equation}
Partition this matrix into the two delays and the four real gain parameters as:
\begin{equation}
\bm J_{\xi,n}
=
\begin{bmatrix}
\bm J_{\tau\tau,n} & \bm J_{\tau\gamma,n}\\
\bm J_{\gamma\tau,n} & \bm J_{\gamma\gamma,n}
\end{bmatrix}
\label{eq:joint_two_path_partition}
\end{equation}
Eliminating the two unknown complex gains by the Schur complement gives the equivalent delay information and covariance
\begin{equation}
\begin{aligned}
\bm J_{\tau,n}^{\mathrm e}
=
\bm J_{\tau\tau,n}
-
\bm J_{\tau\gamma,n}
\bm J_{\gamma\gamma,n}^{\dagger}
\bm J_{\gamma\tau,n},
\quad
\bm C_{\tau,n}
=
(\bm J_{\tau,n}^{\mathrm e})^{-1}
\end{aligned}
\label{eq:joint_two_path_schur}
\end{equation}
provided that the two delays are locally identifiable; otherwise, the corresponding excess-delay bound is unbounded. Since $\Delta\tau_n=\tau_n-\tau_{\mathrm{LOS},n}$, with $\bm q=[-1,\,1]^T$, the variance injected into both position FIMs is:
\begin{equation}
\boxed{
\begin{aligned}
\sigma_{\Delta\tau,n}^{2}
&=\bm q^T\bm C_{\tau,n}\bm q\\
&=[\bm C_{\tau,n}]_{1,1}
 +[\bm C_{\tau,n}]_{2,2}
 -2[\bm C_{\tau,n}]_{1,2}.
\end{aligned}}
\label{eq:tdoa_var}
\end{equation}
Thus, the LOS and target-path powers are treated separately and their delay-estimation covariance is retained rather than set to zero. In the resolved-path limit, where delay gating and/or distinct spatial signatures make the two paths approximately orthogonal, \eqref{eq:tdoa_var} reduces to the useful check
\begin{equation}
\sigma_{\Delta\tau,n}^{2}
\approx
\mathrm{CRLB}_{\mathrm{tar},n}(\tau_n)
+
\mathrm{CRLB}_{\mathrm{LOS},n}(\tau_{\mathrm{LOS},n}),
\label{eq:tdoa_resolved_approx}
\end{equation}
where, for $q\in\{\mathrm{tar},\mathrm{LOS}\}$, the isolated-path bound is obtained from the nuisance-eliminated single-path information in \eqref{eq:single_path_gain_schur}. Because Doppler is compensated in the two-path delay model, it is the conditional bound:
\begin{equation}
\mathrm{CRLB}_{q,n}(\tau_q\mid f_D)
=
\frac{1}{
2\,\mathrm{SNR}_{q,\mathrm{eff},n}N_{R,n}
\sum_{k,m}
(2\pi\widetilde f_{k,m,n})^2
}
\label{eq:isolated_path_delay_crlb}
\end{equation}
The implementation evaluates the full joint FIM in \eqref{eq:joint_two_path_fim}--\eqref{eq:joint_two_path_schur}; \eqref{eq:tdoa_resolved_approx} is used only as a resolved-path consistency check. Likewise, the azimuth and elevation variances are obtained from the angular block of $(\bm F_{\mathrm s}^{\mathrm e})^{-1}$ after eliminating the unknown target coefficient. Under the centered URA and centered active-set coordinates, \eqref{eq:single_path_gain_orthogonality} applies, and the nuisance-eliminated bounds reduce to the active-set forms of Section~\ref{sec:boundary_analysis}. By reusing \eqref{eq:norm_gphi} and \eqref{eq:norm_gtheta}, the receive-array derivative norms of pair $n$ can be written in generalized form as in \eqref{eq:crn_array_const} and \eqref{eq:g_norm_square_general}. Therefore,
\vspace{4pt}
\begin{equation}
\begin{aligned}
&\sigma_{\varphi,n}^{2}
\triangleq
\mathrm{CRLB}(\varphi_n), \quad \sigma_{\vartheta,n}^{2}
\triangleq
\mathrm{CRLB}(\vartheta_n),
\\
&\sigma_{\varphi,n}^{2}
=
\frac{\|\bm g_{\vartheta,n}\|^{2}}
{2N_{\mathcal A,n}\,\mathrm{SNR}_{\mathrm{eff},n}\!\left[\|\bm g_{\varphi,n}\|^{2}\|\bm g_{\vartheta,n}\|^{2}-\Re\!\left\{\bm g_{\varphi,n}^{H}\bm g_{\vartheta,n}\right\}^{2}\right]}\\
&\sigma_{\vartheta,n}^{2}
=
\frac{\|\bm g_{\varphi,n}\|^{2}}
{2N_{\mathcal A,n}\,\mathrm{SNR}_{\mathrm{eff},n}\!\left[\|\bm g_{\varphi,n}\|^{2}\|\bm g_{\vartheta,n}\|^{2}-\Re\!\left\{\bm g_{\varphi,n}^{H}\bm g_{\vartheta,n}\right\}^{2}\right]}
\end{aligned}
\label{eq:angle_var_mode_general}
\end{equation}
For the square half-wavelength URAs used in this paper, $\Re\!\left\{\bm g_{\varphi,n}^{H}\bm g_{\vartheta,n}\right\}=0$, and therefore the azimuth and elevation variances can be injected separately into the position FIM. For a symmetric $3\times3$ position FIM of the form in \eqref{eq:generic_fim_3d}, the corresponding open-form PEB is given by \eqref{eq:peb_open_3d}.
\vspace{4pt}
\begin{equation}
\bm I
=
\begin{bmatrix}
I_{xx} & I_{xy} & I_{xz}\\
I_{xy} & I_{yy} & I_{yz}\\
I_{xz} & I_{yz} & I_{zz}
\end{bmatrix}
\label{eq:generic_fim_3d}
\end{equation}

Hence, the main task is to derive the algorithm-dependent position FIMs for the considered sensing modes and then map them to the 3D PEB through \eqref{eq:peb_open_3d}. The detailed entry-wise open-form expressions are collected in \eqref{eq:Ixx_ta_open}--\eqref{eq:Iyz_d_open}.

\subsubsection{Time-Angle Based Algorithm FIM and PEB}
\label{subsubsec:peb_time_angle}
\vspace{8pt}
In the algorithm, each pair contributes a delay measurement together with AoA azimuth--elevation measurements. Thus, for sensing mode $m\in\{\mathrm{TT},\mathrm{TU},\mathrm{H}\}$, the FIM is in \eqref{eq:fim_ta_mode_open}. Here, $\bm j_n(\bm p)$ is given by \eqref{eq:jacobian_tdoa}, the excess-delay variance $\sigma_{\Delta\tau,n}^{2}$ follows from the joint two-path covariance in \eqref{eq:tdoa_var}, and the angle variances $\sigma_{\varphi,n}^{2}$ and $\sigma_{\vartheta,n}^{2}$ follow from \eqref{eq:angle_var_mode_general}. By grouping the six unique entries, \eqref{eq:fim_ta_mode_open} can be written as:
\begin{equation}
\bm I_{\mathrm{TA},m}(\bm p)
=
\begin{bmatrix}
I_{xx}^{\mathrm{TA},m} & I_{xy}^{\mathrm{TA},m} & I_{xz}^{\mathrm{TA},m}\\
I_{xy}^{\mathrm{TA},m} & I_{yy}^{\mathrm{TA},m} & I_{yz}^{\mathrm{TA},m}\\
I_{xz}^{\mathrm{TA},m} & I_{yz}^{\mathrm{TA},m} & I_{zz}^{\mathrm{TA},m}
\end{bmatrix}
\label{eq:fim_ta_entries_matrix}
\end{equation}

Where the detailed open-form expressions of $I_{xx}^{\mathrm{TA},m}$--$I_{yz}^{\mathrm{TA},m}$ are reported in \eqref{eq:Ixx_ta_open}--\eqref{eq:Iyz_ta_open}. Therefore, the time--angle PEB for mode $m$ becomes:
\begin{equation}
\mathrm{PEB}_{\mathrm{TA},m}
=
\mathrm{PEB}\!\left(\bm I_{\mathrm{TA},m}(\bm p)\right),
\label{eq:peb_time_angle_closed}
\end{equation}
This is obtained by substituting the entries of \eqref{eq:fim_ta_entries_matrix} into the generic mapping in \eqref{eq:peb_open_3d}. Hence, \eqref{eq:peb_time_angle_closed} together with \eqref{eq:Ixx_ta_open}--\eqref{eq:Iyz_ta_open} explicitly shows the dependence on the per-pair sensing SNR, the delay-resolution term $(\Delta f,M,K)$, and the receive-array geometry through \eqref{eq:crn_array_const}--\eqref{eq:angle_var_mode_general}.

\vspace{-18pt}
\subsubsection{Delay Based Algorithm FIM and PEB}
\label{subsubsec:peb_tdoa}
The algorithm uses only excess-delay measurements, and therefore its position FIM contains only the delay term induced by the Jacobian in \eqref{eq:jacobian_tdoa}. For sensing mode $m\in\{\mathrm{TT},\mathrm{TU},\mathrm{H}\}$, the corresponding FIM is given in \eqref{eq:fim_delay_mode_open}. Grouping its six unique entries, matrix written as: \eqref{eq:fim_delay_entries_matrix}.
\begin{equation}
\bm I_{\mathrm{D},m}(\bm p)
=
\sum_{n\in\mathcal{V}_{m}^{\mathrm{D}}}
\frac{1}{c^2\sigma_{\Delta\tau,n}^{2}}\,\bm j_n(\bm p)\bm j_n^{T}(\bm p)
\label{eq:fim_delay_mode_open}
\end{equation}
\begin{equation}
\bm I_{\mathrm{D},m}(\bm p)
=
\begin{bmatrix}
I_{xx}^{\mathrm{D},m} & I_{xy}^{\mathrm{D},m} & I_{xz}^{\mathrm{D},m}\\
I_{xy}^{\mathrm{D},m} & I_{yy}^{\mathrm{D},m} & I_{yz}^{\mathrm{D},m}\\
I_{xz}^{\mathrm{D},m} & I_{yz}^{\mathrm{D},m} & I_{zz}^{\mathrm{D},m}
\end{bmatrix}
\label{eq:fim_delay_entries_matrix}
\end{equation}
where the detailed open-form expressions of $I_{xx}^{\mathrm{D},m}$--$I_{yz}^{\mathrm{D},m}$ are reported in \eqref{eq:Ixx_d_open}--\eqref{eq:Iyz_d_open}. Hence, the delay-only PEB:
\begin{equation}
\mathrm{PEB}_{\mathrm{D},m}
=
\mathrm{PEB}\!\left(\bm I_{\mathrm{D},m}(\bm p)\right)
\label{eq:peb_tdoa_closed}
\end{equation}
which is obtained by substituting the entries of \eqref{eq:fim_delay_entries_matrix} into \eqref{eq:peb_open_3d}. Therefore, \eqref{eq:peb_tdoa_closed} together with \eqref{eq:Ixx_d_open}--\eqref{eq:Iyz_d_open} makes explicit the dependence on the bistatic geometry, the receive-array size $N_{R,n}$, the separate LOS and target powers, and the joint two-path coupling through $\sigma_{\Delta\tau,n}^{2}$ in \eqref{eq:tdoa_var}. A finite value of \eqref{eq:peb_tdoa_closed} requires both locally identifiable per-link delays and a full-rank position FIM $\bm I_{\mathrm{D},m}(\bm p)$, which in turn requires sufficiently diverse, non-coplanar bistatic pairs.

\vspace{-2pt}
\section{Simulation Environment}\label{sec:sim_env}
\vspace{6pt}
\subsection{Indoor factory layout and node deployment}\label{subsec:layout}

Consider a $300 \times 150 \times 10$~m 3D indoor factory hall featuring $18$ TRPs on a $6\times3$ grid ($50$~m spacing, $25$~m wall offset) and $20$ constant UAV targets at $z_{\mathrm{ST}}=8$~m.

\begin{figure}[ht]
    \centering
    \includegraphics[width=\linewidth, height=0.75\linewidth]{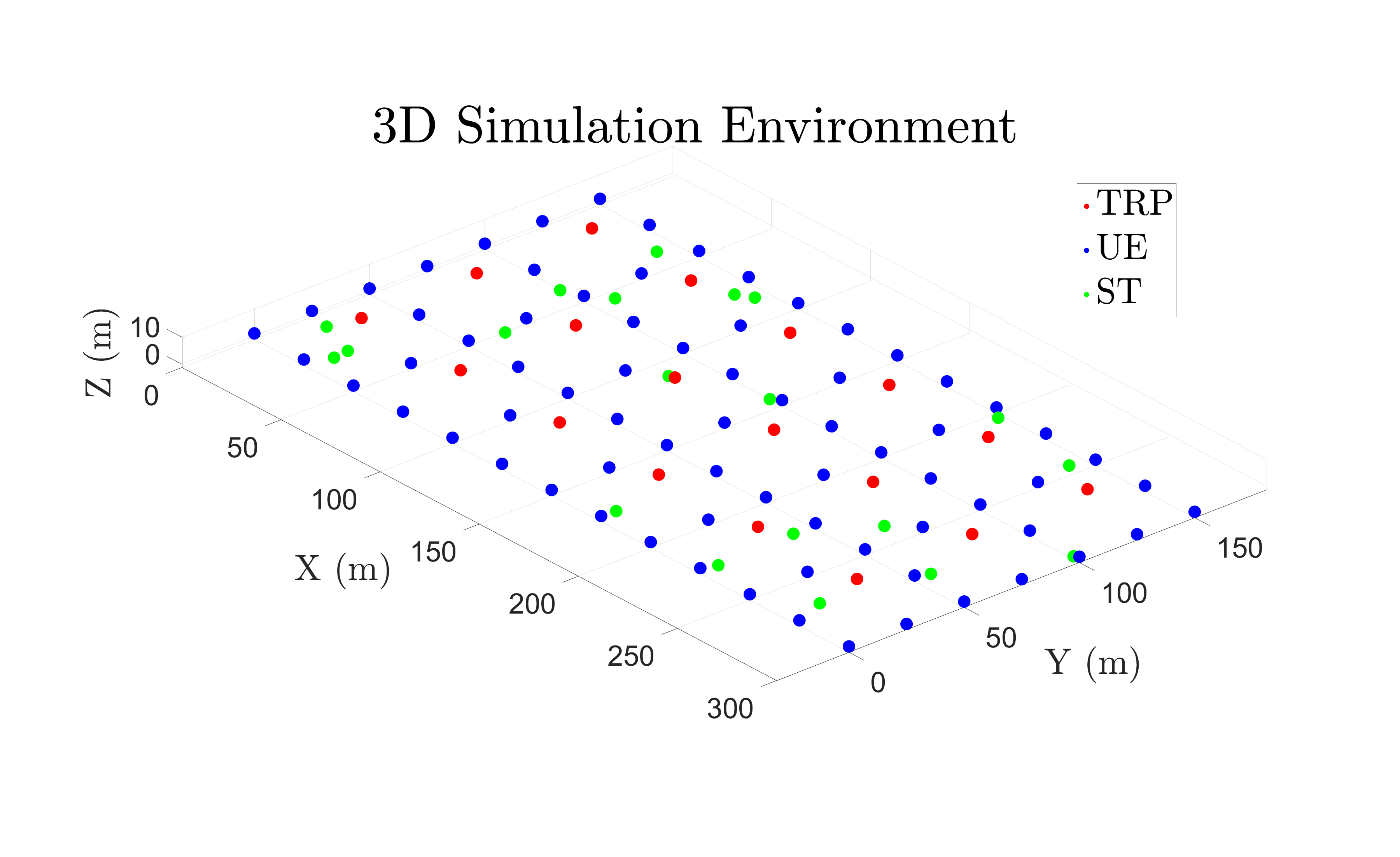}
    \caption{The indoor-factory deployment features three bistatic sensing modes (TRP--TRP, TRP--UE, and Hybrid) using 18 TRPs, 20 active UEs, and 20 target locations, as detailed in Table~\ref{tab:sim_params}.}
    \label{TRP_UE}
\end{figure}

In TRP--UE mode, $20$ active UEs act as sensing receivers with $2\!\times\!2$ URAs alongside TRPs using $8\!\times\!8$ URAs, operating at $f_c=35$~GHz with $0$~dBi gains, $\vert{}\beta\vert{}^2=1$, $T_0=290$~K, and a $9$~dB noise figure. The evaluation considers white AWGN under a single-target Monte-Carlo procedure for each bistatic pair (comprising direct Tx--Rx and target), omitting multipath, clutter, and multi-target interference for Figs.~\ref{fig:err_vs_snr}--\ref{fig:cdf}.

\vspace{-0.5cm}
\begin{figure}[hb]
    \centering
    \includegraphics[width=\linewidth, height=0.75\linewidth]{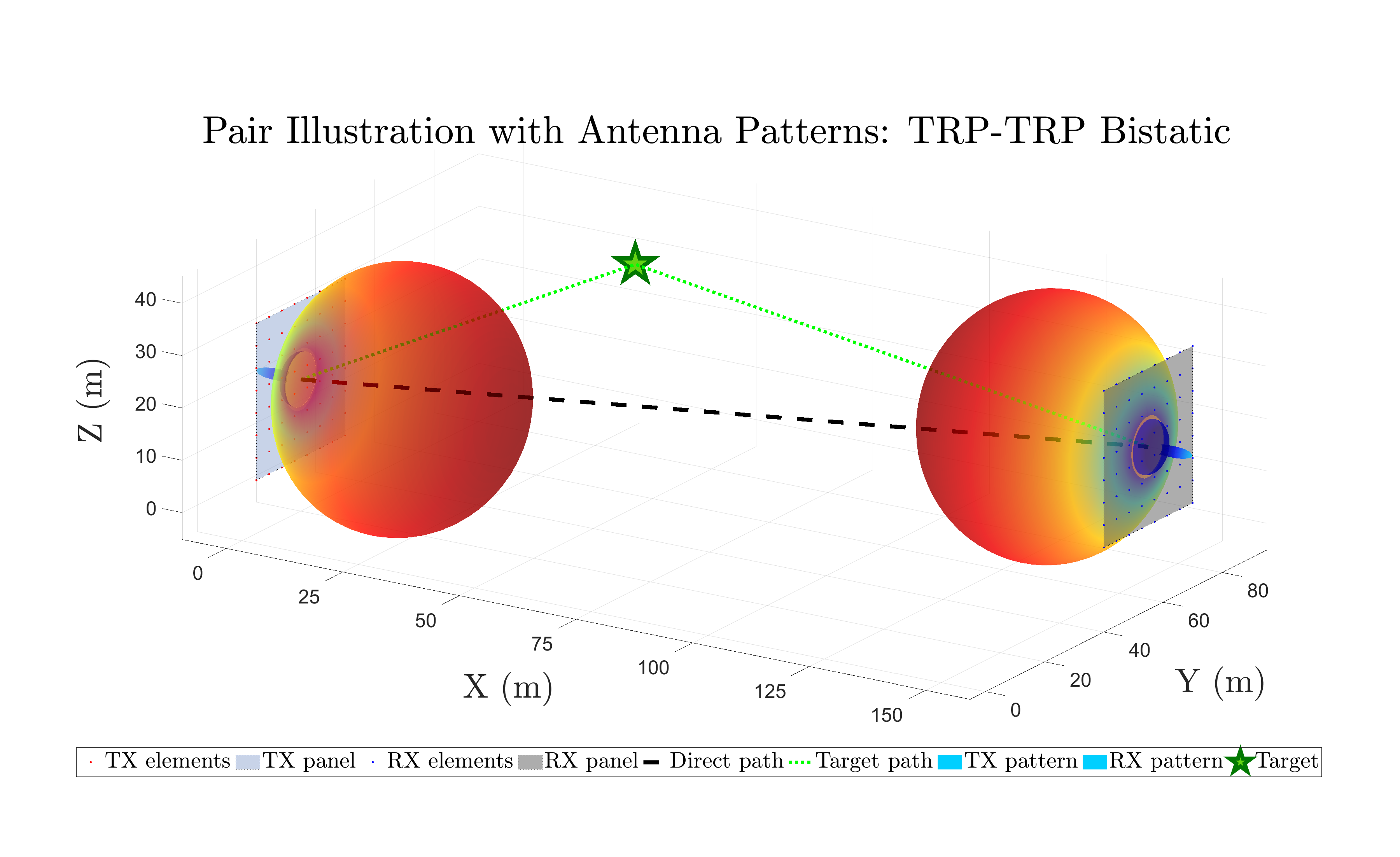}
    \caption{3D llustration of one bistatic sensing pair with antennas.}
    \label{Pair}
\end{figure}

\vspace{-12pt}
\subsection{Bistatic sensing modes, links, and sensing pairs}\label{subsec:modes}

Three network-assisted bistatic sensing modes are evaluated. \textbf{TRP--TRP bistatic sensing:} a TRP transmits and a (different) TRP receives the reflected echo from the ST.
\textbf{TRP--UE bistatic sensing:} a TRP transmits and a UE receives the reflected echo from the ST.
\textbf{Hybrid sensing:} the sensing function fuses heterogeneous bistatic measurements collected from both TRP--TRP and TRP--UE modes to improve positioning robustness and accuracy.

A \emph{link} is any propagation segment between nodes in $\{\mathrm{TRP}, \mathrm{UE}, \mathrm{ST}\}$. A \emph{sensing pair} is a transmitter--target--receiver configuration used for positioning and comprises three links: $\mathrm{TRP}\!\rightarrow\!\mathrm{ST}$, $\mathrm{ST}\!\rightarrow\!r$, and the baseline direct path $\mathrm{TRP}\!\rightarrow\!r$, where $r\in\{\mathrm{TRP},\mathrm{UE}\}$. Accordingly, TRP--TRP uses a TRP receiver, TRP--UE uses a UE receiver, and the Hybrid case combines both. Practical pair scheduling is driven by a common coarse 3D target estimate from an initial acquisition or previous tracking occasion rather than by the true target position. For each target/Monte-Carlo realization, the estimate is generated as $\widehat{\bm p}_{\mathrm{sch}}=\bm p+\bm e$, with independent horizontal standard deviation $\sigma_{xy}=5$~m and vertical standard deviation $\sigma_z=2$~m, and is projected onto the known factory volume. 

All three coordinates are therefore uncertain, and the fixed $8$~m target height is not supplied to the scheduler. The same $\widehat{\bm p}_{\mathrm{sch}}$ is used by TRP--UE, TRP--TRP, and Hybrid sensing. Candidate links are ranked by their estimated bistatic path lengths. Each mode uses exactly $N_{\mathrm{pair}}$ selected links under the same per-pair waveform, active-resource-element allocation, and transmit-power rule; hence, for a given waveform, the post-acquisition sensing budget is $N_{\mathrm{pair}}N_{\mathcal A}$ active-resource-element observations in every mode, while common coarse-acquisition input contributes identical overhead to modes. 

Hybrid sensing ranks the TRP--UE and TRP--TRP candidates separately and applies a fixed near-$50/50$ selected-link quota instead of searching their union. For odd $N_{\mathrm{pair}}$, the extra link is alternated between the two pair types over target/trial realizations. The reported numerical curves use $N_{\mathrm{pair}}=10$.

\subsection{Bandwidth, delay resolution, and angular resolution}\label{subsec:resolution}
\vspace{4pt}
The noise variance at one receive-chain FFT output is $\sigma_{N,\mathrm{RE}}^2=k_{\mathrm B}T_0F\Delta f$. If OFDM symbol $m$ contains $K_{\mathrm a,m}$ active tones, its summed FFT-bin noise bandwidth is
\begin{equation}
B_{\mathrm{occ},m}=K_{\mathrm a,m}\Delta f.
\label{eq:occupied_noise_bandwidth}
\end{equation}
This occupied-tone measure must be distinguished from the frequency aperture that controls delay resolution. For an arbitrary active set $\mathcal A$, the relevant delay-information quantity is the second central frequency moment $S_{ff}$ in \eqref{eq:fim_summ}. A convenient span measure is
\begin{equation}
B_{\mathrm{span}}
\triangleq
f_{\max,\mathcal A}-f_{\min,\mathcal A}+\Delta f
\label{eq:frequency_span}
\end{equation}

For a contiguous allocation, $B_{\mathrm{occ}}\approx B_{\mathrm{span}}\approx K\Delta f$, and the usual delay-resolution approximation $\Delta\tau\approx1/B_{\mathrm{span}}$ applies. For a PRS comb, only one out of every $K_{\mathrm{comb}}$ tones is active in a given symbol, but adjacent active PRS tones are separated by $K_{\mathrm{comb}}\Delta f$ and may span almost the full allocated PRB range. Therefore, replacing the PRS allocation by $K$ contiguous tones with spacing $\Delta f$ gives an incorrect $S_{ff}$; the exact mapping in \eqref{eq:prs_active_set} is used instead.

In NR-style parametrization, a carrier grid of $N_{\mathrm{RB}}$ resource blocks contains $12N_{\mathrm{RB}}$ subcarriers. The communication-reuse waveform activates the full 106-RB grid, corresponding to 1272 subcarriers and $N_{\mathcal A}=1272M$ active REs over $M$ reconstructed OFDM symbols. The PRS resource uses 104 RBs with RB offset 1, 12 PRS symbols starting from OFDM symbol 0, comb size 2, RE offset 0, and $N_{\mathrm{ID}}^{\mathrm{PRS}}=0$. Consequently, each PRS symbol contains $12\times104/2=624$ active tones and the slot contains $N_{\mathcal A}=7488$ active PRS resource elements. The allocated PRS frequency span is approximately $12\times104\Delta f$; at $\Delta f=240$~kHz, it equals $299.52$~MHz, whereas the full communication grid spans $305.28$~MHz. The corresponding monostatic range-resolution scaling is $\Delta R\approx\frac{c}{2B_{\mathrm{span}}}$ while the exact bistatic delay and position bounds are evaluated through $S_{ff}$ and the geometry-dependent position FIM. Waveform comparisons use the same average power per active RE; they are not equal-total-energy comparisons across allocations with different active-tone counts. The angular resolving power is governed by the array aperture and follows the qualitative scaling $\Delta\theta\propto\frac{\lambda}{D_{\mathrm{ap}}}$, where $D_{\mathrm{ap}}$ is the effective aperture dimension.
\subsection{Target reflectivity and radar cross section (RCS)}\label{subsec:rcs}
\vspace{6pt}
Radar cross section (RCS) $\sigma$ is an effective area (in m$^2$) that quantifies the strength of the backscattered echo.
It is often expressed in dBsm via:
\begin{equation}
\sigma_{\text{dBsm}} = 10\log_{10}\!\big(\sigma\,[\text{m}^2]\big), \
\sigma\,[\text{m}^2] = 10^{\sigma_{\text{dBsm}}/10}
\end{equation}

\paragraph{Reference-SNR convention and physical link budget.}
To generate controlled SNR sweeps without suppressing the physical distance and RCS dependence, horizontal axis is defined as reference bistatic-echo SNR and power:
\begin{equation}
G_{\mathrm{bi,ref}}
=
\frac{G_TG_R\lambda^2\sigma_{\mathrm{ref}}}
{(4\pi)^3d_{T,\mathrm{ref}}^2d_{R,\mathrm{ref}}^2},
\label{eq:reference_bistatic_gain}
\end{equation}
with $d_{T,\mathrm{ref}}=d_{R,\mathrm{ref}}=30$~m and $\sigma_{\mathrm{ref}}=1$~m$^2$ ($0$~dBsm). The Monte-Carlo implementation uses the fixed transmit-array factor $|\beta|^2=1$ for both the direct and target components; the general analytical model retains $|\beta|^2$ explicitly. For a requested per-resource-element reference SNR, the per-RE transmit power:
\begin{equation}
P_{\mathrm{RE}}
=
\frac{\mathrm{SNR}_{\mathrm{ref,pre}}\,\sigma_{N,\mathrm{RE}}^2}
{G_{\mathrm{bi,ref}}\,\mathbb{E}\{|x[k,m]|^2\}}.
\label{eq:reference_transmit_power}
\end{equation}
When the plotted SNR is specified after coherent reciprocal processing, the simulator first removes the corresponding active-resource-element processing gain to obtain $\mathrm{SNR}_{\mathrm{ref,pre}}$.The same $P_{\mathrm{RE}}$ and $\sigma_{N,\mathrm{RE}}^2$ are then used for the actual direct and target paths, whose complex amplitudes are $\sqrt{P_{\mathrm{RE}}G_{\mathrm{LOS},n}}$ and $\sqrt{P_{\mathrm{RE}}G_{\mathrm{bi},n}}$, respectively. Thus, the RCS sweep is in physical dBsm and changes received target power linearly with $\sigma_{\mathrm{RCS}}$; no independently tuned target-to-LOS power ratio is used.

For canonical scatterers, RCS can be related to physical area; e.g., for a perfectly-conducting flat plate at normal incidence, $\sigma$ scales with the square of plate area and inversely with $\lambda^2$ (illustrating the strong dependence on both target size and carrier frequency). In the literature, measured human RCS values depend on frequency, aspect angle, motion, and clothing. As an example at mmWave automotive bands, the frequency-and-azimuth averaged RCS of a standing pedestrian has been reported on the order of a few negative dBsm (i.e., a few tenths of a square meter). Small UAV/quadcopter RCS values are typically smaller (more negative dBsm), depending on size, material and frequency.

\subsection{Key simulation parameters}\label{subsec:params}
\vspace{6pt}
Table~\ref{tab:sim_params} provides a compact summary of the principal quantitative settings. The exact PRS mapping, link-budget convention, receiver-noise model, robust-fusion constants, and scheduling rule are specified in the corresponding signal, algorithm, and simulation descriptions.

\begin{table}[ht]
\centering
\caption{Summary of principal quantitative settings.}
\label{tab:sim_params}
\footnotesize
\renewcommand{\arraystretch}{1.05}
\begin{tabular}{@{}>{\raggedright\arraybackslash}p{0.25\columnwidth} >{\raggedright\arraybackslash}p{0.69\columnwidth}@{}}
\hline
\textbf{Group} & \textbf{Setting} \\
\hline
Scenario & InF hall $300\times150\times10$~m; 18 TRPs, 20 UEs, 20 independent UAV targets @ $z_{\mathrm{ST}}=8$~m \\
Carrier/arrays & $f_c=35$~GHz; $8\times8$ TRP and $2\times2$ UE \\
Waveforms & PRS (104 RBs, 12 symbols, comb 2); OFDM reuse (BPSK/16QAM/64QAM on 106 RBs) \\
Sweeps & $\mathrm{SNR}_{\mathrm{ref}}=0{:}5{:}30$~dB; SCS $\in\{30,60,120,240\}$~kHz; RCS $\in\{-10,-5,0\}$~dBsm \\
Scheduling & Noisy 3D estimate; 10 links per mode\\
\hline
\end{tabular}
\end{table}
\vspace{-0.25cm}
\section{Simulation Results}
\label{sec:sim_results}
\vspace{8pt}
\subsection{Qualitative Signal-Domain Illustrations}
\label{subsec:sim_measurements}

\begin{figure}[ht]
    \centering
    \includegraphics[width=1\linewidth, height=0.85\linewidth]{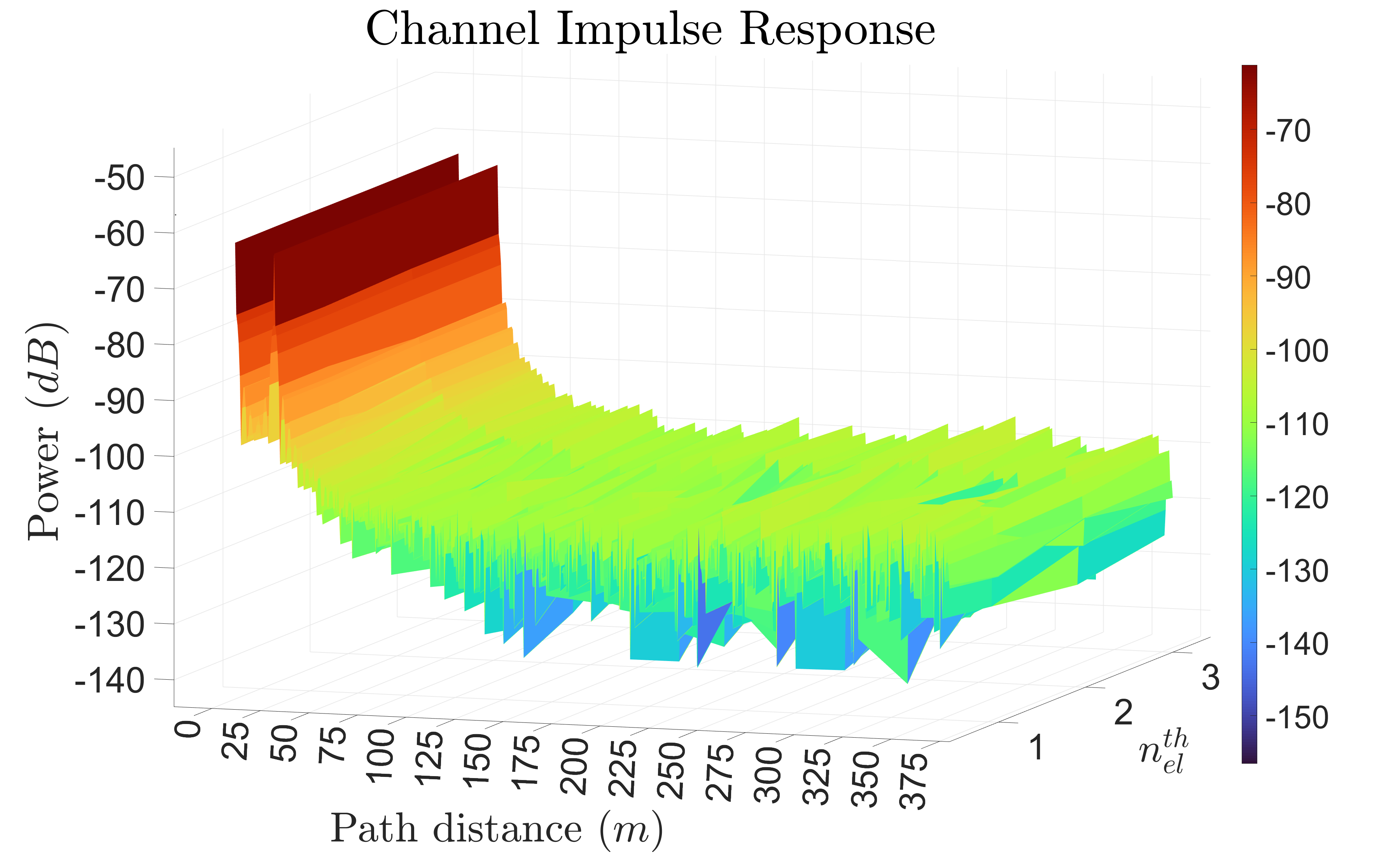}
    \caption{Qualitative 3D delay-domain power illustration. The horizontal axis is path length (m), the second axis is receiver-antenna index, and the color bar is power (dB). The early coherent ridge is the direct Tx--Rx component and the delayed ridge is the ST echo; any additional peaks are illustrative only and are not used in the quantitative PEB/positioning results.}
    \label{fig:ch_3d}
\end{figure}

\begin{figure}[hb]
    \centering
    \includegraphics[width=1\linewidth, height=0.7\linewidth]{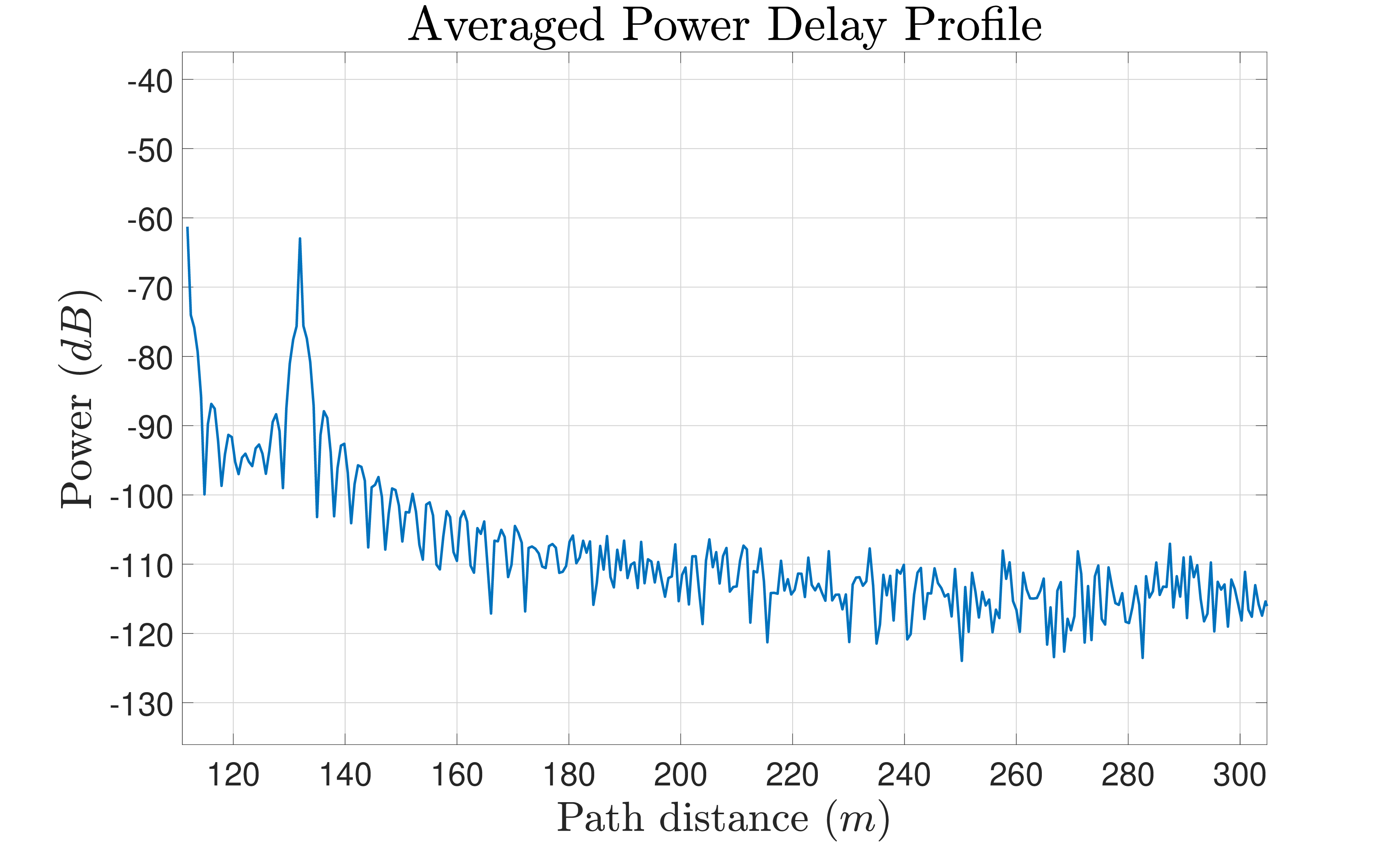}
    \caption{Qualitative averaged power-delay profile.}
    \label{fig:pdp}
\end{figure}

\begin{figure}[ht]
    \centering
    \includegraphics[width=1\linewidth, height=0.65\linewidth]{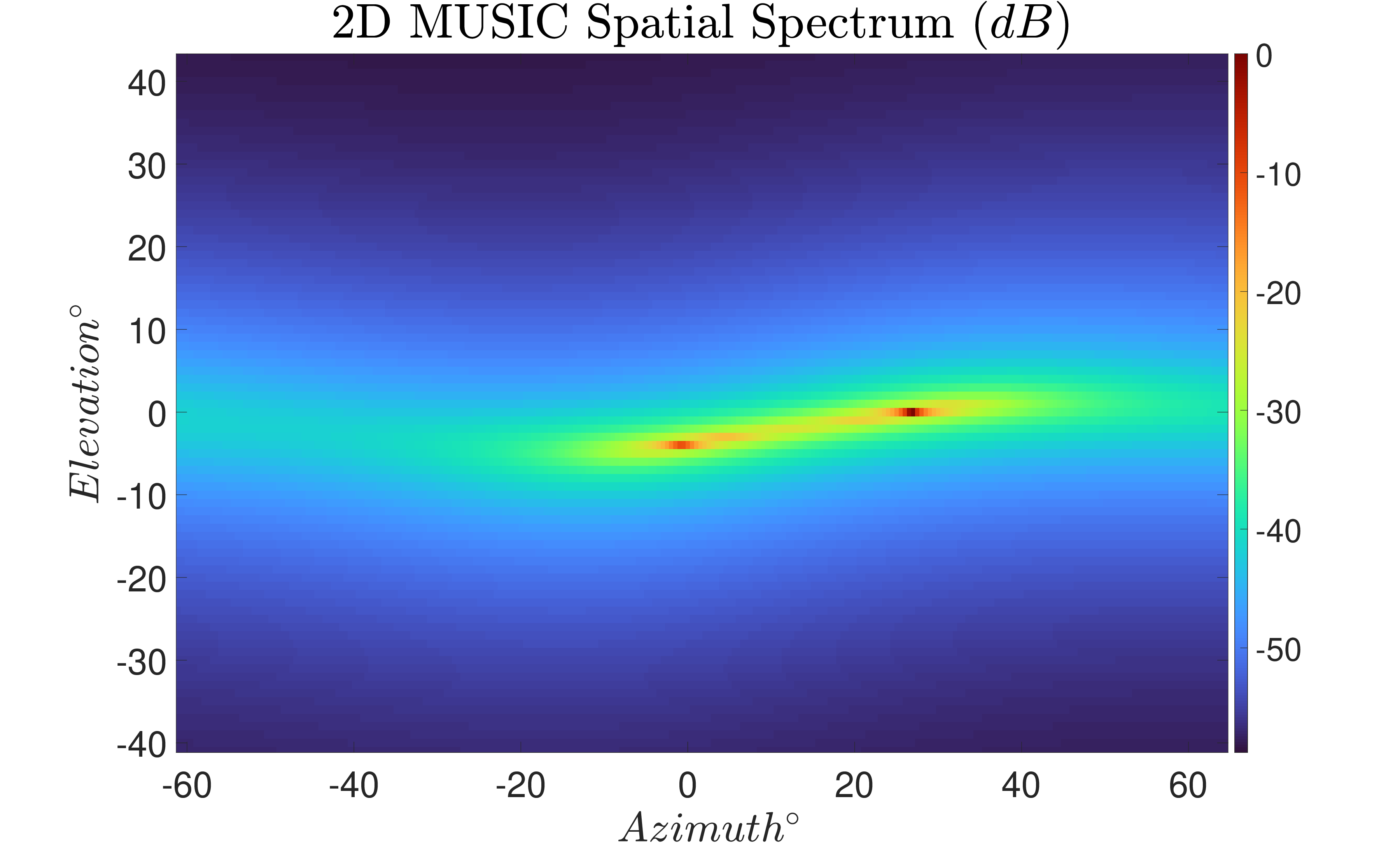}
    \caption{Qualitative 2D MUSIC spectrum for ST DoA estimation. The panel is a zoom of the full search grid; axes are azimuth/elevation.}
    \label{fig:music_2d}
\end{figure}

Figures~\ref{fig:ch_3d}--\ref{fig:music_2d} are qualitative signal-domain illustrations used to explain the delay and angle extraction stages. They are not the channel realizations used to generate the quantitative positioning curves or PEBs in Figs.~\ref{fig:err_vs_snr}--\ref{fig:cdf}; those results use the common direct-path-plus-target-echo model stated in Section~\ref{subsec:layout}. Figure~\ref{fig:ch_3d} illustrates received delay-domain power across the receiver antenna elements. The horizontal axis is expressed as path length in meters through $d=c\tau$, the second axis is the receiver-antenna index, and the color bar represents power in dB. 

The dominant early ridge corresponds to the direct baseline path, while the delayed ridge corresponds to the bistatic ST echo with path length approximately $r_T+r_R$. Any additional weaker peaks shown in this illustrative snapshot represent possible unmodeled propagation components and are included only to motivate delay gating and spatial processing; they are not included in the quantitative PEB comparison. The spatial coherence of the direct and target components across antenna elements supports DoA estimation, while their separation in path length supports excess-delay estimation.

Figure~\ref{fig:pdp} shows the corresponding qualitative averaged power-delay profile (PDP). The horizontal axis is path length in meters and the vertical axis is normalized power in dB. The early peak identifies the direct path and the later peak identifies the bistatic ST return. Any remaining fluctuations are interpreted only as illustrative residual components and noise. The separation between the two principal peaks motivates the direct-path cancellation and target-peak search used by the excess-delay estimator; larger frequency aperture improves this resolvability and reduces the delay bound.

\vspace{-0.5cm}

\subsubsection{2D DoA estimation using 2D MUSIC Algorithm}
\vspace{4pt}
To estimate the direction of arrival of the target echo, we employ 2D MUSIC over azimuth $\varphi$ and elevation $\vartheta$ using the URA steering-vector model in Section~II-B. Specifically, the receiver forms the sample spatial covariance matrix as in \eqref{eq:cov_doa}, computes the noise subspace via eigendecomposition, and evaluates the 2D MUSIC pseudo-spectrum \eqref{eq:music_2d} on a discrete azimuth--elevation grid. The numerical search uses azimuth angles from $-180^{\circ}$ to $180^{\circ}$ and elevation angles from $-90^{\circ}$ to $90^{\circ}$ with $1^{\circ}$ spacing. Fig.~\ref{fig:music_2d} shows the zoomed region $\varphi\in[-60^{\circ},60^{\circ}]$, $\vartheta\in[-40^{\circ},40^{\circ}]$ of that search, where brighter regions indicate candidate arrival directions.

TRPs use an $8\times 8$ URA while UEs use a compact $2\times 2$ URA. Consequently, TRP-side DoA estimation provides a narrower mainlobe and higher angular resolution, whereas UE-side DoA estimation is more angle-limited and yields broader peaks. If an illustrative spectrum contains multiple local maxima, delay-domain gating associates the selected angular peak with the ST echo. Overall, Figs.~\ref{fig:ch_3d}--\ref{fig:music_2d} explain the intermediate delay and DoA measurements, while the quantitative results reported next remain tied to single-target two-path model.

\subsection{Quantitative Positioning Results}
\vspace{4pt}

\begin{figure}[H]
    \centering
    \includegraphics[width=1\linewidth, height=0.85\linewidth]{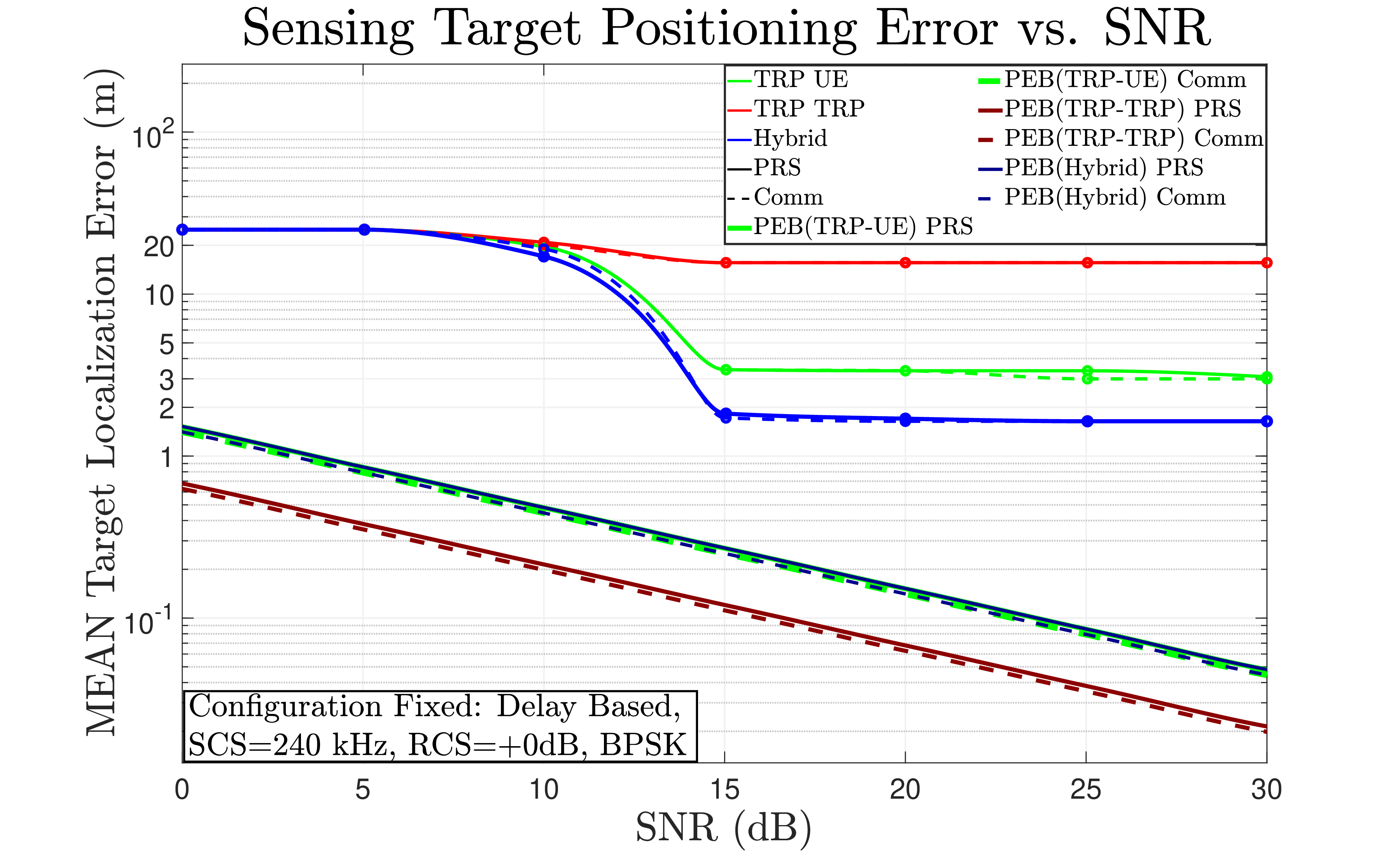}
    \caption{$40$~m-penalized mean 3D positioning error (m) versus reference bistatic-echo SNR (dB) for the delay-based algorithm with fixed SCS $=240$~kHz, $\sigma_{\mathrm{RCS}}=0$~dBsm, and BPSK signaling. Solid curves denote PRS, dashed curves denote communication waveform reuse, and colors indicate sensing modes.}
    \label{fig:err_vs_snr}
\end{figure}

\begin{figure}[H]
    \centering
    \includegraphics[width=1\linewidth,  height=0.75\linewidth]{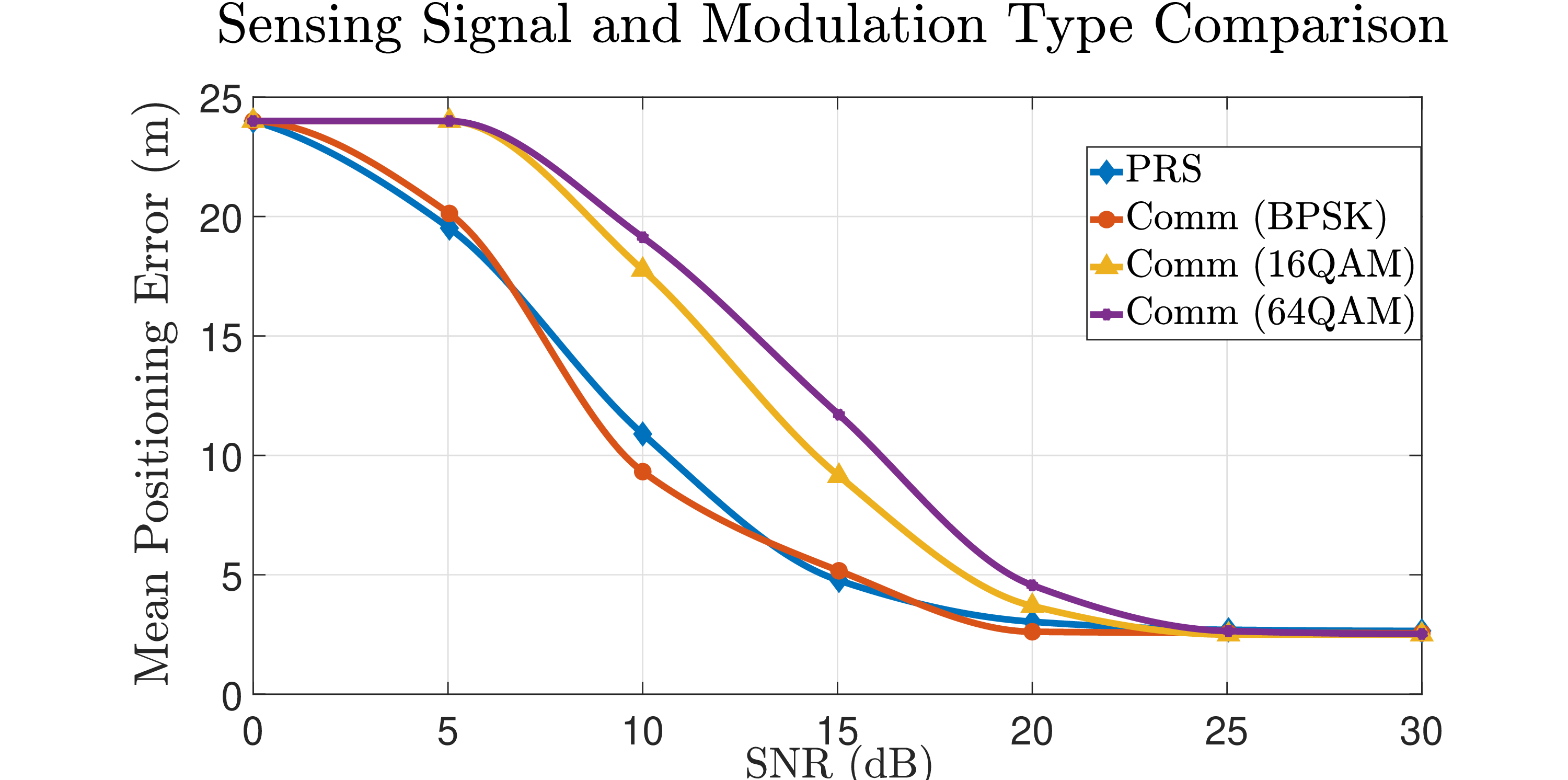}
    \caption{Hybrid-mode $40$~m-penalized mean 3D positioning error (m) versus reference bistatic-echo SNR (dB) for the delay-based algorithm at $\Delta f=240$~kHz and $\sigma_{\mathrm{RCS}}=0$~dBsm, comparing PRS with communication waveform reuse using BPSK, 16QAM, and 64QAM under the adopted unweighted reciprocal-normalization receiver.}
    \label{fig:mod_sweep}
\end{figure}

\begin{figure}[h]
    \centering
    \includegraphics[width=1\linewidth,  height=0.75\linewidth]{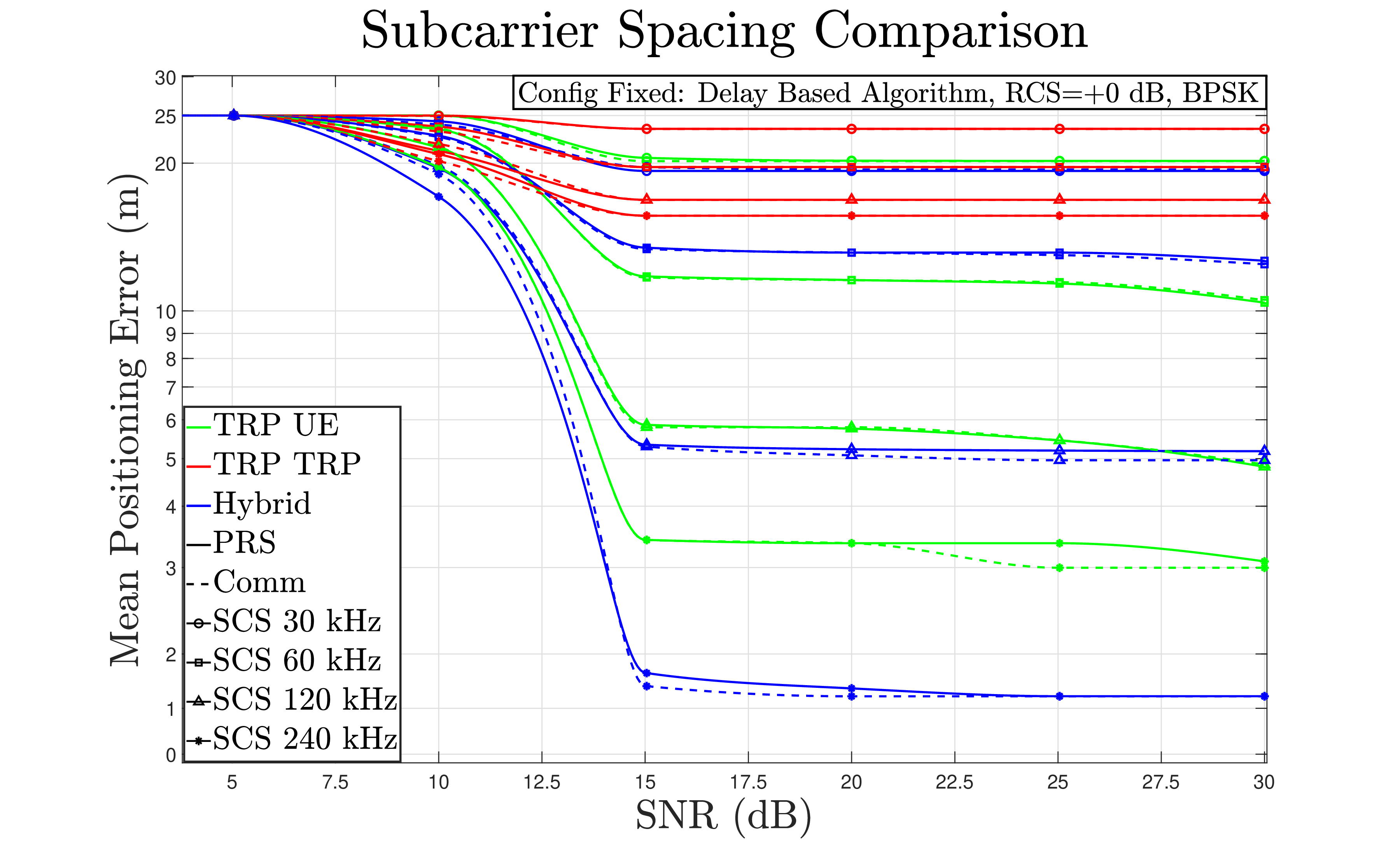}
    \caption{Mean 3D positioning error (m) versus reference bistatic-echo SNR (dB) for SCS values $\{30,60,120,240\}$~kHz.}
    \label{fig:scs_sweep}
\end{figure}

\begin{figure}[ht]
    \centering
    \includegraphics[width=1\linewidth,  height=0.7\linewidth]{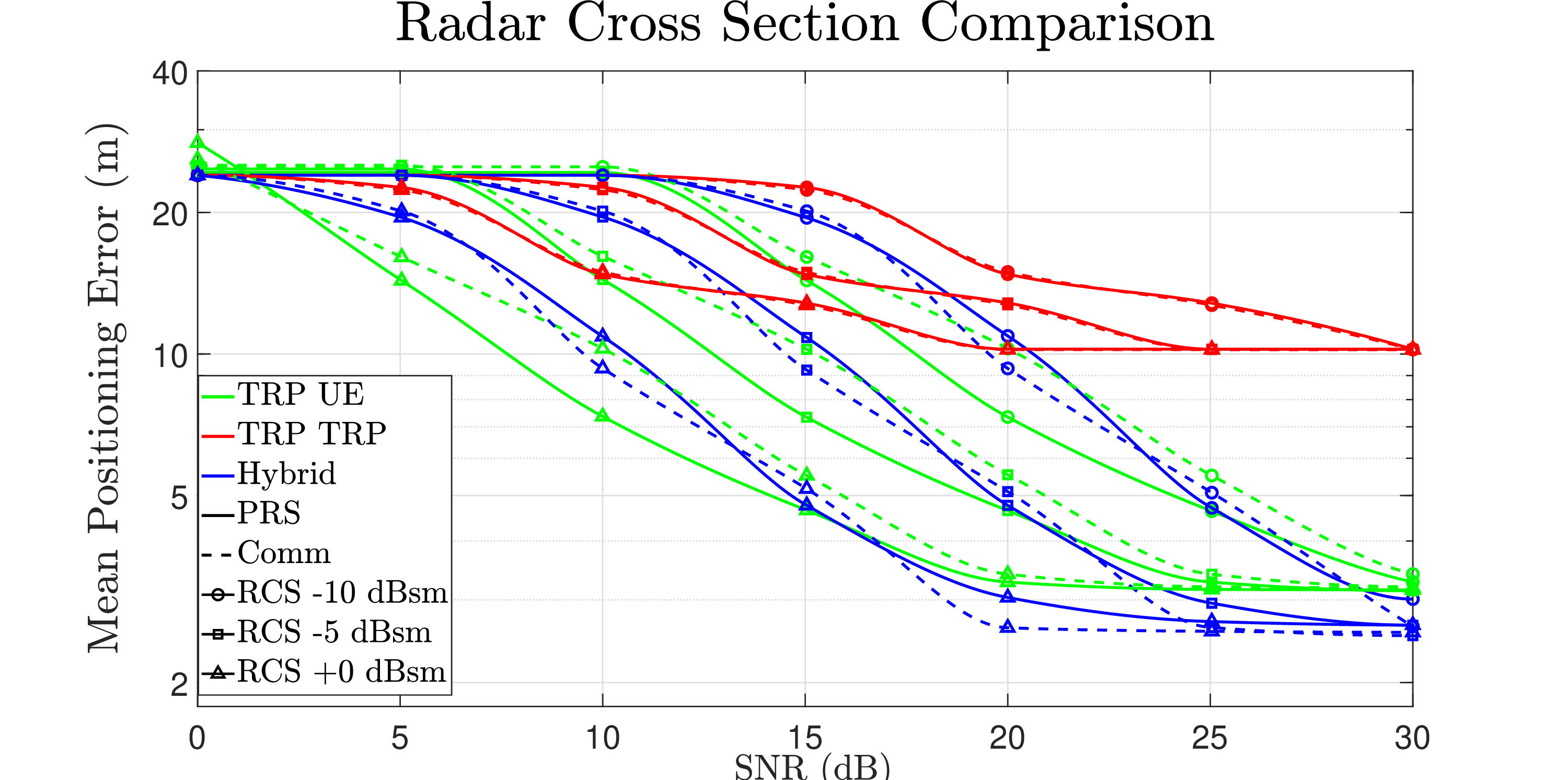}
    \caption{$40$~m-penalized mean 3D positioning error (m) vs. bistatic-echo SNR (dB) for RCS $\{-10,-5,0\}$~dBsm at SCS $=240$~kHz.}
    \label{fig:rcs_sweep}
\end{figure}

\begin{figure}[hb]
    \centering
    \includegraphics[width=1\linewidth,  height=0.7\linewidth]{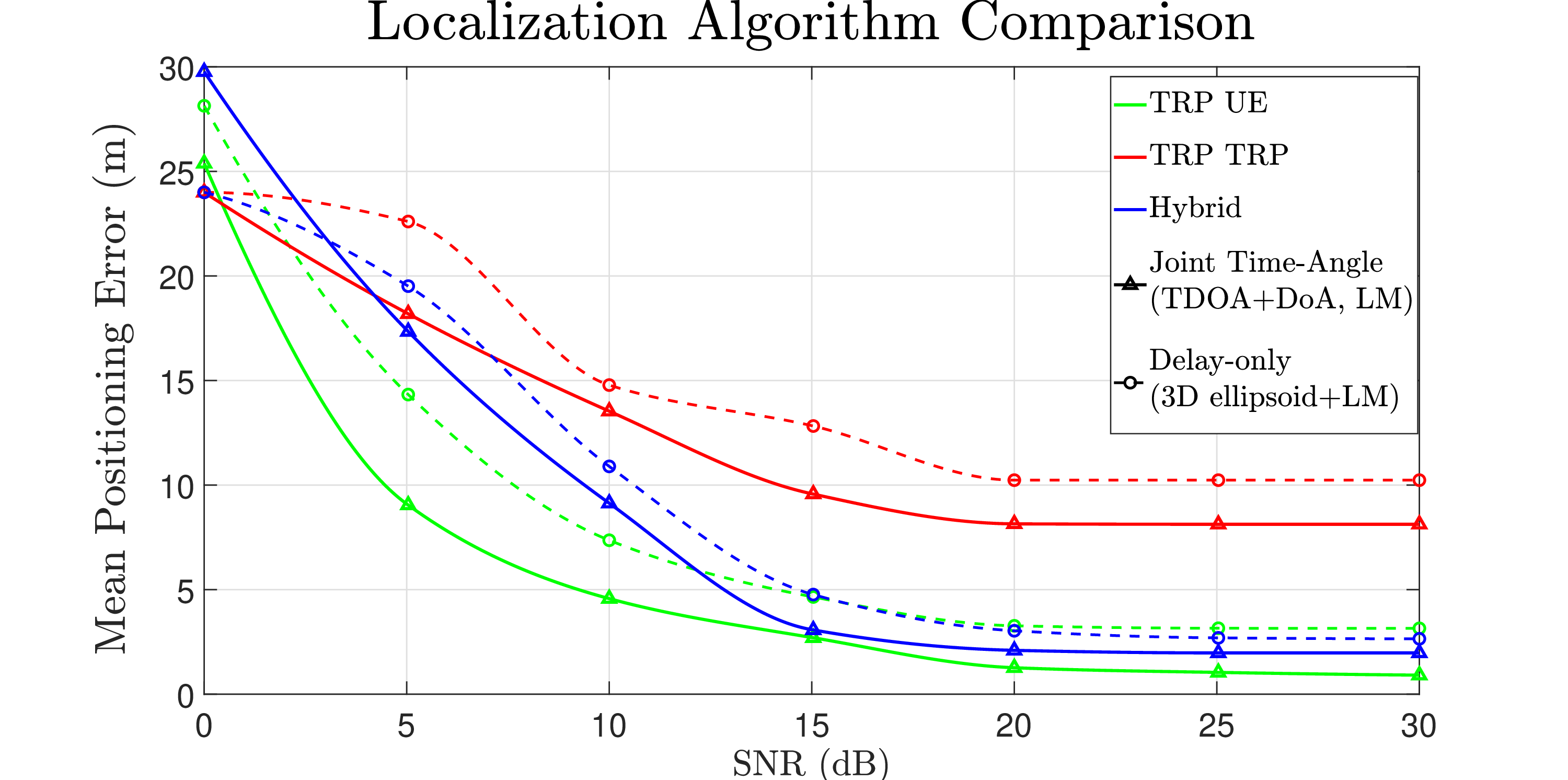}
    \caption{3D positioning error (m) vs reference bistatic-echo SNR (dB) for the robust variance-weighted joint multi-link Time--Angle and delay-only algorithms under PRS signaling with SCS $=240$~kHz and $\sigma_{\mathrm{RCS}}=0$~dBsm. Colors denote the sensing mode; triangle and circle markers denote the Time--Angle and delay-only algorithms, respectively.}
    \label{fig:alg}
\end{figure}

Quantitative positioning performance is evaluated with the same single-target two-path observation used by the PEB: one direct Tx--Rx component, one ST-reflected component, and spatially white AWGN. The 20 stationary target locations are processed independently. Results are reported for TRP--UE, TRP--TRP, and Hybrid sensing. Figures~\ref{fig:err_vs_snr}--\ref{fig:rcs_sweep} use the delay-based 3D ellipsoid multilateration algorithm to isolate waveform, bandwidth, and RCS effects. Figure~\ref{fig:alg} compares the delay-based and Time--Angle algorithms under a common baseline, while Fig.~\ref{fig:cdf} gives conditional empirical CDFs of successful delay-based estimates. 

Every PRS-dependent curve and PEB in Figs.~\ref{fig:err_vs_snr}--\ref{fig:cdf} uses the explicit 104-RB, comb-2 mapping described in Section~\ref{subsec:resolution}; the Monte-Carlo receiver, delay-profile point-spread function, and theoretical FIM use the same active resource-element mask. Figures~\ref{fig:err_vs_snr}--\ref{fig:alg} retain the previously adopted mean-error convention for continuity: a non-finite position estimate contributes a $40$~m penalty. Figure~\ref{fig:cdf} excludes non-finite estimates; $P_{\mathrm{fail}}$ is defined separately in \eqref{eq:failure_probability} and is not inferred from the conditional CDF. Positioning error is defined as the Euclidean 3D distance $\|\widehat{\bm p}-\bm p\|$ between the estimated and true ST positions. Unless stated otherwise, the curves use $N_{\mathrm{pair}}=10$ selected sensing pairs per mode and $250$ outer realizations over the same 20 target locations, yielding $5{,}000$ positioning attempts per SNR. 

The oracle closest-pair benchmark has been removed. A mode-independent coarse acquisition estimate is generated for each target/Monte-Carlo realization as $\widehat{\bm p}_{\mathrm{sch}}=\bm p+\bm e$, where $e_x$ and $e_y$ are independent zero-mean Gaussian variables with standard deviation $5$~m and $e_z$ has standard deviation $2$~m. The estimate is projected onto the known hall volume; neither the true target coordinates nor the fixed $8$~m height are supplied to the scheduler. The same estimate realization is used by every sensing mode. TRP--UE and TRP--TRP candidates are ranked by estimated bistatic path length. 

Every mode uses exactly $N_{\mathrm{pair}}$ links with the same waveform/resource-element configuration and per-resource-element transmit-power rule, giving the same post-acquisition budget of $N_{\mathrm{pair}}N_{\mathcal A}$ active-resource-element observations. 
Hybrid sensing ranks the two pair types separately and divides the selected-link budget equally for $N_{\mathrm{pair}}=10$; the general odd-budget implementation alternates the extra link between pair types. 

The residual-delay search interval and nonlinear position refinement use only the known factory volume. Figure~\ref{fig:cdf} uses 1,000 outer realizations over 20 target locations, yielding $20{,}000$ attempts per waveform. Every active received resource element is generated as $\bm y[k,m]=\bm h[k,m]x[k,m]+\bm n[k,m]$ and divided by the actual known symbol $x[k,m]$, so 16QAM and 64QAM retain the exact symbol-dependent post-normalization noise. The horizontal axis is the reference bistatic-echo SNR defined in \eqref{eq:reference_bistatic_gain}--\eqref{eq:reference_transmit_power}. The direct and target amplitudes use the same $P_{\mathrm{RE}}$ through $\sqrt{P_{\mathrm{RE}}G_{\mathrm{LOS},n}}$ and $\sqrt{P_{\mathrm{RE}}G_{\mathrm{bi},n}}$, and the RCS values are physical dBsm values. Any in-plot shorthand ``RCS$=0$ dB'' denotes $0$~dBsm.

Figure~\ref{fig:err_vs_snr} shows the baseline reference-SNR sweep for the delay-based algorithm with fixed SCS $=240$~kHz, $\sigma_{\mathrm{RCS}}=0$~dBsm, and BPSK signaling. Solid curves correspond to PRS and dashed curves to communication waveform reuse. Within this panel, Hybrid enters the low-error regime earliest and has the smallest high-SNR floor, TRP--UE is intermediate, and TRP--TRP remains substantially higher.

The TRP--TRP behavior is attributed to \emph{unfavorable geometry}, i.e., poor conditioning of the delay-only position FIM/Jacobian rather than merely long link distance: the selected TRP--TRP ellipsoids can have similar baselines and weak intersection angles because the TRPs lie on a regular infrastructure grid, so their range-sum gradients are less diverse. 

UE-assisted and Hybrid modes add heterogeneous receiver viewpoints, which improve the rank and conditioning of the position FIM. An important observation is that the PRS and BPSK-reuse curves are nearly coincident for all three sensing modes. This is consistent with the constant-modulus nature of BPSK, for which reciprocal normalization does not create symbol-dependent noise enhancement. The overlaid delay PEBs use the joint two-path construction in \eqref{eq:joint_two_path_fim}--\eqref{eq:tdoa_var}; hence, the strong direct path is not assigned the target-echo delay variance, and any LOS--target delay coupling is retained. Their PRS and BPSK-reuse versions remain very close because both signals are constant modulus under the adopted reciprocal receiver. Numerical implementation also forms the augmented single-path FIM in \eqref{eq:single_path_fim_partition} and applies \eqref{eq:single_path_gain_schur}. 

The delay--Doppler block is formed from the exact moments $S_{ff}$, $S_{tt}$, and $S_{ft}$ of each waveform. Thus, PRS uses its actual comb spacing and symbol-dependent offsets, whereas the communication waveform uses its full-grid active set. The absolute high-SNR values in this figure are specific to the stated carrier, bandwidth, RCS, array sizes, node layout, and scheduler. They are therefore not presented as a direct state-of-the-art record; scope-normalized literature discussion in Section~I.

Figure~\ref{fig:mod_sweep} focuses on the Hybrid mode and compares PRS with communication waveform reuse employing BPSK, 16QAM, and 64QAM at fixed SCS $=240$~kHz and $\sigma_{\mathrm{RCS}}=0$~dBsm. The Monte-Carlo implementation performs $\bm z[k,m]=\bm y[k,m]/x[k,m]$ on every active resource element after adding pre-normalization AWGN. Therefore, the 16QAM and 64QAM curves include the actual symbol-by-symbol heteroscedastic noise. 

PRS and BPSK-reuse almost overlap, whereas 16QAM and especially 64QAM are shifted to the right and require larger SNR to reach the same error level. This shift is a receiver-dependent robustness cost of the adopted unweighted reciprocal-normalization processor in the transition region; it is not claimed to be a fundamental penalty of communication-waveform reuse or of higher-order QAM itself.

For readability, Figs.~\ref{fig:scs_sweep} and~\ref{fig:rcs_sweep} use the axis limits shown in each panel; in both cases, non-finite estimates still contribute the same $40$~m penalty. Figure~\ref{fig:scs_sweep} studies the subcarrier-spacing sweep with $\Delta f\in\{30,60,120,240\}$~kHz at $\sigma_{\mathrm{RCS}}=0$~dBsm and BPSK signaling. Since a larger SCS scales the physical separation of the active frequencies and therefore increases $S_{ff}$ and the frequency span, the delay resolution improves significantly, which appears in the figure as both an earlier threshold transition and a lower high-SNR error floor. For PRS, this scaling is computed from the mapped comb tones in every PRS symbol; it is not obtained by treating the 624 active tones as a contiguous block. This improvement is most pronounced for Hybrid and TRP--UE sensing, where the $240$~kHz case reaches low-meter accuracy while the $30$~kHz case remains clearly bandwidth-limited even at high SNR. TRP--TRP also benefits from larger bandwidth, but its final error floor stays much higher than the others, showing that geometry rather than delay resolution alone becomes the dominant bottleneck in that mode. For every SCS value, the PRS and BPSK-reuse curves remain very close, again confirming that constant-envelope waveform reuse is highly attractive for sensing.

Figure~\ref{fig:rcs_sweep} examines the influence of target strength by sweeping $\sigma_{\mathrm{RCS}}\in\{-10,-5,0\}$~dBsm at fixed SCS $=240$~kHz with BPSK signaling. As expected, stronger targets shift the performance transition to lower SNR, whereas weaker targets require several additional decibels before the error collapses to its low-error regime. For the considered layout, array sizes, scalar gains, and equal-budget scheduler, the relative ordering remains unchanged: Hybrid is the most robust, TRP--UE is second, and TRP--TRP is the most fragile. In particular, Hybrid and TRP--UE still converge to low-meter errors once the SNR becomes sufficient, while TRP--TRP saturates at a much larger error even for the high-RCS target. Once again, the gap between PRS and BPSK-reuse is small, which confirms that constant-envelope waveform reuse can preserve most of the positioning performance of PRS in the delay-based pipeline.

Figure~\ref{fig:alg} directly compares the two positioning algorithms under a common PRS baseline with SCS $=240$~kHz and $\sigma_{\mathrm{RCS}}=0$~dBsm. Both methods jointly process the same selected links under the same sensing-resource budget, factory-volume constraint, and multi-start candidate family. The Time--Angle method obtains $\widehat{\bm p}_{\mathrm g}$ from a joint range-sum/DoA-ray fit and then evaluates all delay and angular uncertainty values at that common estimate before the robust covariance-weighted refinement in \eqref{eq:ta_joint_objective}. In the delay-only method, the Time--Angle candidate is used only as an optional starting point; all optimization and candidate selection remain based on delay residuals. 

The range residual is normalized by the standard deviation derived from the joint LOS--target excess-delay variance, while the DoA is represented as a tangent-plane lateral residual whose covariance reflects the plug-in link SNR and exact active resource-element set. Bounded variance ratios, block-Huber weighting, and the measured-data geometric safeguard in \eqref{eq:ta_geometric_guard} prevent a single locally overconfident or sidelobe-contaminated DoA estimate from driving the solution to a spurious boundary point. The delay-only method continues to fit the bistatic range-sum residuals only. Thus, the comparison no longer contrasts simple averaging with joint nonlinear optimization; both pipelines perform joint multi-link fusion, while each uses a residual model appropriate to its measurements. 

The resulting separation between the curves should be interpreted through measurement information, receiver angular resolution, robust weighting, and sensing geometry. An inherent superiority of delay-only positioning is not consequently claimed. The delay-only estimator is retained as the main reference in Figs.~\ref{fig:err_vs_snr}--\ref{fig:rcs_sweep} to isolate waveform, bandwidth, and RCS effects with one common positioning pipeline.

Figure~\ref{fig:cdf} reports the empirical CDF of the positioning error conditioned on successful localization for the delay-based algorithm at $\mathrm{SNR}_{\mathrm{ref}}=30$~dB, SCS $=240$~kHz, and $\sigma_{\mathrm{RCS}}=-10$~dBsm, comparing PRS with 64QAM communication-waveform reuse. Failed estimates are excluded from the CDF rather than replaced by the $40$~m penalty. In the legend, ``Comm'' denotes 64QAM reuse; $P_{\mathrm{fail}}$ is defined separately in \eqref{eq:failure_probability} and is not encoded by the conditional curves. The $0.5$ crossing gives the conditional median, while the conditional RMSE is defined over the same finite-error samples in \eqref{eq:conditional_rmse}. The Hybrid and TRP--UE curves are steeper and more left-shifted than the TRP--TRP curves for the adopted configuration, indicating a lower and tighter successful-estimate error distribution. TRP--TRP exhibits the heaviest tail. The 64QAM-reuse curves are shifted to the right of their PRS counterparts, consistently with the symbol-dependent noise enhancement of the adopted unweighted reciprocal-normalization receiver.

\vspace{-1cm}
\subsubsection{Dependence on hardware and deployment}
\label{subsubsec:hardware_deployment_dependence}

\begin{figure}[ht!]
    \centering
    \includegraphics[width=1\linewidth,  height=0.75\linewidth]{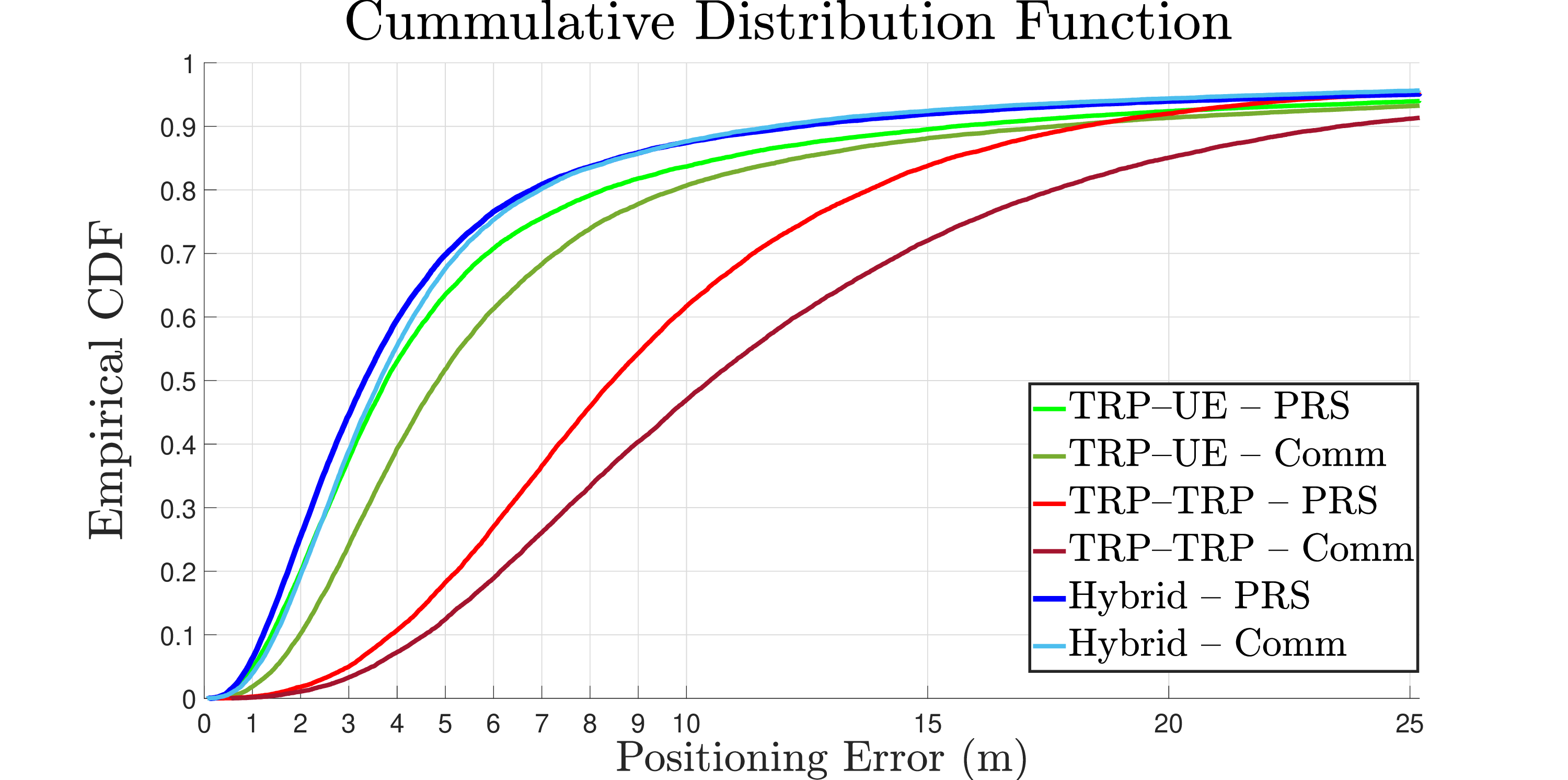}
    \caption{Conditional empirical CDF of successful 3D position estimates (m) for the delay-based algorithm at $\mathrm{SNR}_{\mathrm{ref}}=30$~dB, SCS $=240$~kHz, and $\sigma_{\mathrm{RCS}}=-10$~dBsm, comparing PRS with 64QAM communication-waveform reuse. Failed estimates are excluded; ``Comm'' denotes 64QAM reuse, and $P_{\mathrm{fail}}$ is defined in \eqref{eq:failure_probability}.}
    \label{fig:cdf}
\end{figure}

The ordering observed in Figs.~\ref{fig:err_vs_snr}--\ref{fig:cdf} is conditional on the adopted $2\!\times\!2$ UE and $8\!\times\!8$ TRP receive arrays, half-wavelength element spacing, scalar gains $G_T=G_R=1$, 18-TRP layout, target set, and equal selected-link budget $N_{\mathrm{pair}}=10$. It is not a general ordering between TRP--UE and TRP--TRP sensing. The angular information in \eqref{eq:fim_phiphi}--\eqref{eq:fim_phitheta} is proportional to the steering-vector derivative norms in \eqref{eq:norm_gphi}--\eqref{eq:norm_gtheta}; these norms increase with receive-element count and with the square of the physical element spacing over the non-aliased range. A larger UE or TRP aperture can therefore reduce the Time--Angle PEB and change the relative value of the two receiver types. The link SNRs in \eqref{eq:snr_eff_pair}--\eqref{eq:snr_eff_los_pair} also scale with the scalar gains and the physical path lengths, so stronger TRP receive gain can narrow the baseline TRP--UE/TRP--TRP gap.

TRP density and placement affect the delay-only result through the position FIM in \eqref{eq:fim_delay_mode_open}, which is a sum of outer products of the bistatic range-gradient vectors. Additional nodes improve the bound only when the scheduled pairs add sufficiently distinct gradient directions; a clustered or highly regular layout can remain poorly conditioned despite a larger node count. Spatially distributed TRPs, favorable baseline orientations, or alternative pair scheduling can therefore improve TRP--TRP performance and may alter the baseline ordering. 

\subsubsection{Failure-aware interpretation}
\label{subsubsec:failure_aware_interpretation}

Let $\mathcal{S}$ denote the set of trials that return a finite 3D position and let $N_{\mathrm{tot}}$ be the total number of attempts. The failure probability, conditional median and RMSE:
\begin{equation}
P_{\mathrm{fail}}=1-\frac{|\mathcal{S}|}{N_{\mathrm{tot}}},
\label{eq:failure_probability}
\end{equation}
\begin{equation}
e_{\mathrm{med}|\mathcal{S}}=\operatorname{median}\!\left\{\|\widehat{\bm p}_i-\bm p_i\|:i\in\mathcal{S}\right\},
\label{eq:conditional_median}
\end{equation}
\begin{equation}
e_{\mathrm{RMSE}|\mathcal{S}}=\sqrt{\frac{1}{|\mathcal{S}|}\sum_{i\in\mathcal{S}}\|\widehat{\bm p}_i-\bm p_i\|^2}.
\label{eq:conditional_rmse}
\end{equation}
Neither \eqref{eq:conditional_median} nor \eqref{eq:conditional_rmse} replaces a failure by $40$~m. Figures~\ref{fig:err_vs_snr}--\ref{fig:alg} are explicitly labelled as $40$~m-penalized means, whereas Fig.~\ref{fig:cdf} uses only finite estimates. Its $0.5$ crossing gives the conditional median, \eqref{eq:conditional_rmse} defines the RMSE over the same successful-estimate set, and $P_{\mathrm{fail}}$ must be evaluated separately from \eqref{eq:failure_probability}. Reliability and successful-estimate accuracy must therefore be interpreted jointly; no conclusion is drawn from the penalty-based mean alone. Figures~\ref{fig:err_vs_snr}--\ref{fig:cdf} support the following conclusions for the adopted deployment. Hybrid sensing gives the lowest delay-based penalized-mean and conditional-CDF errors, followed by TRP--UE, while TRP--TRP is the most challenging configuration because of the scheduled geometry. This baseline performance ordering is non-generalizable due to configuration-specific array aperture, node density, and geometry. Algorithm behavior is driven by available delay/angle data and resolution rather than optimization differences. While PRS and waveform reuse perform similarly for constant-modulus signals, QAM requires higher SNR. Finally, larger frequency apertures and stronger target returns reduce positioning errors, with $P_{\mathrm{fail}}$ remaining an independent reliability metric.

\section{Conclusion}
\label{sec:conclusion}
\vspace{6pt}

\subsection{Summary of Contributions and Key Findings}
\vspace{2pt}
This paper investigated UE-assisted passive target positioning for cellular ISAC, with a focus on hybrid operation that fuses TRP--TRP and TRP--UE bistatic measurements. A unified OFDM-based signal, channel, and noise model was developed for PRS-based sensing and communication waveform reuse. The PRS sequence sample, PRB allocation, OFDM-symbol index, comb size, configured and symbol-dependent offsets, and normalization are linked explicitly through the active resource-element mapping. The model distinguishes the bistatic power gain from the complex amplitude, keeps per-RE/total transmit power, scalar antenna gains, RCS, propagation distances, array factors, and thermal noise explicit, and uses an unweighted reciprocal-normalization receiver with exact symbol-dependent noise and a clearly identified average-noise approximation for closed-form analysis. Based on this model, the real and imaginary parts of the unknown target-reflection coefficient were included in the augmented FIM and eliminated through a Schur complement before deriving the equivalent CRLBs for delay, Doppler, and angular parameters. These bounds use the exact active frequency--time moments, so the PRS comb spacing, symbol-dependent offsets, and any nonzero delay--Doppler cross moment are retained. These bounds were mapped to algorithm-specific positioning error bounds (PEBs) for two representative positioning pipelines: (i) a robust variance-weighted joint multi-link Time--Angle approach combining TDOA and DoA, and (ii) a delay-only 3D ellipsoid multilateration approach based on excess-delay/TDOA measurements. For the latter mapping, each LOS--target pair is evaluated through a joint two-path FIM with distinct physical powers, nuisance complex gains eliminated by a Schur complement, and the resulting delay covariance retained. This unified treatment establishes a consistent analytical link between waveform design, sensing geometry, and achievable positioning performance.

Monte-Carlo results in a realistic indoor-factory deployment highlight several important insights. All sensing modes are compared using the same practical coarse 3D acquisition estimate and identical selected-pair and sensing-resource budgets, while Hybrid sensing ranks TRP--UE and TRP--TRP candidates separately and uses a fixed near-$50/50$ composition rather than an unrestricted union search. First, UE participation and Hybrid TRP--UE/TRP--TRP fusion significantly improve robustness by increasing bistatic viewpoint diversity and reducing sensitivity to weak or geometrically ill-conditioned links, demonstrating the value of UE-assisted sensing in practical deployments. Second, the revised algorithm comparison uses joint nonlinear multi-link fusion for both Time--Angle and delay-only positioning. The Time--Angle pipeline evaluates all uncertainty values at one common joint geometric estimate, whitens the range and tangent-plane DoA residuals with link-specific delay and lateral covariances, bounds excessive reliability ratios, applies block-Huber weighting, and retains the stable geometric estimate when the weighted candidate fails a measured-data consistency safeguard. Its relative performance is therefore attributed to delay/angle measurement quality, receiver aperture, outlier robustness, and geometry; in particular, compact UE arrays can limit the benefit of DoA information, but the manuscript does not claim that delay-only positioning is inherently superior. Third, increasing the occupied bandwidth (e.g., via larger subcarrier spacing) enhances delay resolution and reduces the achievable error floor, whereas higher-order QAM primarily shifts the break-point SNR for the adopted unweighted reciprocal-normalization receiver, influencing robustness rather than high-SNR accuracy in the considered implementation. Fourth, the analytical dependence of the FIM and PEB on receiver element count, physical aperture, scalar gains, pair geometry, and node placement demonstrates that the observed sensing-mode ordering is configuration specific. Finally, the explicit separation of the $40$~m-penalized mean, conditional CDF, $P_{\mathrm{fail}}$, and penalty-free conditional accuracy definitions prevents failure penalty from being interpreted as a physical localization error.
\vspace{-0.25cm}
\subsection{Implications and Future Directions}
\vspace{8pt}
Overall, the results indicate that hybrid bistatic sensing with UE assistance is a promising approach for practical passive positioning in the considered cellular ISAC deployment, particularly under heterogeneous link quality and varying geometrical conditions. The findings suggest that, for the evaluated array sizes and node layout, gains can be realized by leveraging measurement diversity and hybrid fusion strategies. In particular, reliable joint multi-link fusion and increased effective bandwidth appear to be key enablers for improved positioning accuracy in realistic deployments. These observations further suggest that practical ISAC network designs can benefit from assigning complementary sensing roles to infrastructure nodes and UEs, instead of relying solely on network-side sensing. Hybrid fusion improves the penalized-mean and conditional-error distributions for the evaluated configuration; the analytical dependence on hardware and geometry requires this statement to remain conditional on the tested arrays, gains, scheduler, and deployment.

Future work will focus on enhancing receiver-side angular processing, including higher-resolution DoA estimation techniques and larger effective apertures, as well as developing adaptive sensing-pair selection and fusion mechanisms. In addition, more comprehensive modeling of multipath propagation, clutter effects, and multi-target scenarios will be essential to further bridge the gap between practical estimators and the derived theoretical bounds, thereby advancing the reliability and scalability of cellular ISAC systems in real-world environments. Another relevant direction is the joint design of waveform configuration, sensing-pair scheduling, and estimator selection, so that the network can adapt its positioning strategy to the operating geometry and channel conditions. Experimental validation under practical hardware impairments and synchronization limitations would also be valuable for assessing the real-world feasibility of the proposed hybrid sensing framework.

\section*{Acknowledgment}
\vspace{8pt}
This work was funded and supported in part by the MSCA INTEGRATE and the SNS JU 6G-DISAC Horizon Europe projects under Grant Agreement numbers 101072924 and 101139130, respectively. This project is also funded by UK Research and Innovation (UKRI).

\section*{Appendix: Detailed Open-Form FIM Entries}
\vspace{8pt}
Detailed entry-wise expansions of the algorithm-dependent position FIMs are provided for the two 3D positioning pipelines. Appendix~A expands the Time--Angle FIM in \eqref{eq:fim_ta_mode_open}, used to evaluate $\mathrm{PEB}_{\mathrm{TA},m}$ in \eqref{eq:peb_time_angle_closed}. 

Appendix~B expands the delay-only FIM in \eqref{eq:fim_delay_mode_open}, used to evaluate $\mathrm{PEB}_{\mathrm{D},m}$ in \eqref{eq:peb_tdoa_closed}. The geometric quantities $r_{T,n}$, $r_{R,n}$, $\bm p$, $\bm s_{T,n}$, $\bm s_{R,n}$, and $\bm j_n(\bm p)$ are defined with the positioning model. The excess-delay variance $\sigma_{\Delta\tau,n}^{2}$ is supplied by the joint LOS--target covariance in \eqref{eq:tdoa_var}, while $\sigma_{\varphi,n}^{2}$ and $\sigma_{\vartheta,n}^{2}$ follow from \eqref{eq:angle_var_mode_general}. The expansions make the position-FIM entries explicit for the TRP--TRP, TRP--UE, and Hybrid modes.

\subsection*{A. Detailed Time--Angle FIM Entries}
\begingroup
\small
\allowdisplaybreaks
\setlength{\jot}{2pt}
\begin{align}
I_{xx}^{\mathrm{TA},m}
&=
\sum_{n\in\mathcal{V}_{m}^{\mathrm{TA}}}
\Bigg[
\frac{1}{c^2\sigma_{\Delta\tau,n}^{2}}
\left(
\frac{x-x_{T,n}}{r_{T,n}}
+
\frac{x-x_{R,n}}{r_{R,n}}
\right)^2
\nonumber\\
&\quad
+
\frac{1}{\sigma_{\varphi,n}^{2}}
\frac{(y-y_{R,n})^2}
{\left((x-x_{R,n})^2+(y-y_{R,n})^2\right)^2}
\nonumber\\
&\quad
+
\frac{1}{\sigma_{\vartheta,n}^{2}}
\frac{(x-x_{R,n})^2(z-z_{R,n})^2}
{\left((x-x_{R,n})^2+(y-y_{R,n})^2\right)r_{R,n}^4}
\Bigg],
\label{eq:Ixx_ta_open}\\
I_{yy}^{\mathrm{TA},m}
&=
\sum_{n\in\mathcal{V}_{m}^{\mathrm{TA}}}
\Bigg[
\frac{1}{c^2\sigma_{\Delta\tau,n}^{2}}
\left(
\frac{y-y_{T,n}}{r_{T,n}}
+
\frac{y-y_{R,n}}{r_{R,n}}
\right)^2
\nonumber\\
&\quad
+
\frac{1}{\sigma_{\varphi,n}^{2}}
\frac{(x-x_{R,n})^2}
{\left((x-x_{R,n})^2+(y-y_{R,n})^2\right)^2}
\nonumber\\
&\quad
+
\frac{1}{\sigma_{\vartheta,n}^{2}}
\frac{(y-y_{R,n})^2(z-z_{R,n})^2}
{\left((x-x_{R,n})^2+(y-y_{R,n})^2\right)r_{R,n}^4}
\Bigg],
\label{eq:Iyy_ta_open}\\
I_{zz}^{\mathrm{TA},m}
&=
\sum_{n\in\mathcal{V}_{m}^{\mathrm{TA}}}
\Bigg[
\frac{1}{c^2\sigma_{\Delta\tau,n}^{2}}
\left(
\frac{z-z_{T,n}}{r_{T,n}}
+
\frac{z-z_{R,n}}{r_{R,n}}
\right)^2
\nonumber\\
&\quad
+
\frac{1}{\sigma_{\vartheta,n}^{2}}
\frac{(x-x_{R,n})^2+(y-y_{R,n})^2}{r_{R,n}^4}
\Bigg],
\label{eq:Izz_ta_open}\\
I_{xy}^{\mathrm{TA},m}
&=
\sum_{n\in\mathcal{V}_{m}^{\mathrm{TA}}}
\Bigg[
\frac{1}{c^2\sigma_{\Delta\tau,n}^{2}}
\left(
\frac{x-x_{T,n}}{r_{T,n}}
+
\frac{x-x_{R,n}}{r_{R,n}}
\right)
\nonumber\\
&\quad
\left(
\frac{y-y_{T,n}}{r_{T,n}}
+
\frac{y-y_{R,n}}{r_{R,n}}
\right)
\nonumber\\
&\quad
-
\frac{1}{\sigma_{\varphi,n}^{2}}
\frac{(x-x_{R,n})(y-y_{R,n})}
{\left((x-x_{R,n})^2+(y-y_{R,n})^2\right)^2}
\nonumber\\
&\quad
+
\frac{1}{\sigma_{\vartheta,n}^{2}}
\frac{(x-x_{R,n})(y-y_{R,n})(z-z_{R,n})^2}
{\left((x-x_{R,n})^2+(y-y_{R,n})^2\right)r_{R,n}^4}
\Bigg],
\label{eq:Ixy_ta_open}\\
I_{xz}^{\mathrm{TA},m}
&=
\sum_{n\in\mathcal{V}_{m}^{\mathrm{TA}}}
\Bigg[
\frac{1}{c^2\sigma_{\Delta\tau,n}^{2}}
\left(
\frac{x-x_{T,n}}{r_{T,n}}
+
\frac{x-x_{R,n}}{r_{R,n}}
\right)
\nonumber\\
&\quad
\left(
\frac{z-z_{T,n}}{r_{T,n}}
+
\frac{z-z_{R,n}}{r_{R,n}}
\right)
\nonumber\\
&\quad
-
\frac{1}{\sigma_{\vartheta,n}^{2}}
\frac{(x-x_{R,n})(z-z_{R,n})}{r_{R,n}^4}
\Bigg],
\label{eq:Ixz_ta_open}\\
I_{yz}^{\mathrm{TA},m}
&=
\sum_{n\in\mathcal{V}_{m}^{\mathrm{TA}}}
\Bigg[
\frac{1}{c^2\sigma_{\Delta\tau,n}^{2}}
\left(
\frac{y-y_{T,n}}{r_{T,n}}
+
\frac{y-y_{R,n}}{r_{R,n}}
\right)
\nonumber\\
&\quad
\left(
\frac{z-z_{T,n}}{r_{T,n}}
+
\frac{z-z_{R,n}}{r_{R,n}}
\right)
\nonumber\\
&\quad
-
\frac{1}{\sigma_{\vartheta,n}^{2}}
\frac{(y-y_{R,n})(z-z_{R,n})}{r_{R,n}^4}
\Bigg].
\label{eq:Iyz_ta_open}
\end{align}
\endgroup

\subsection*{B. Detailed Delay-Only FIM Entries}
\begingroup
\small
\allowdisplaybreaks
\setlength{\jot}{2pt}
\begin{align}
I_{xx}^{\mathrm{D},m}
&=
\sum_{n\in\mathcal{V}_{m}^{\mathrm{D}}}
\frac{1}{c^2\sigma_{\Delta\tau,n}^{2}}
\left(
\frac{x-x_{T,n}}{r_{T,n}}
+
\frac{x-x_{R,n}}{r_{R,n}}
\right)^2,
\label{eq:Ixx_d_open}\\
I_{yy}^{\mathrm{D},m}
&=
\sum_{n\in\mathcal{V}_{m}^{\mathrm{D}}}
\frac{1}{c^2\sigma_{\Delta\tau,n}^{2}}
\left(
\frac{y-y_{T,n}}{r_{T,n}}
+
\frac{y-y_{R,n}}{r_{R,n}}
\right)^2,
\label{eq:Iyy_d_open}\\
I_{zz}^{\mathrm{D},m}
&=
\sum_{n\in\mathcal{V}_{m}^{\mathrm{D}}}
\frac{1}{c^2\sigma_{\Delta\tau,n}^{2}}
\left(
\frac{z-z_{T,n}}{r_{T,n}}
+
\frac{z-z_{R,n}}{r_{R,n}}
\right)^2,
\label{eq:Izz_d_open}\\
I_{xy}^{\mathrm{D},m}
&=
\sum_{n\in\mathcal{V}_{m}^{\mathrm{D}}}
\frac{1}{c^2\sigma_{\Delta\tau,n}^{2}}
\left(
\frac{x-x_{T,n}}{r_{T,n}}
+
\frac{x-x_{R,n}}{r_{R,n}}
\right)
\nonumber\\
&\quad
\left(
\frac{y-y_{T,n}}{r_{T,n}}
+
\frac{y-y_{R,n}}{r_{R,n}}
\right),
\label{eq:Ixy_d_open}\\
I_{xz}^{\mathrm{D},m}
&=
\sum_{n\in\mathcal{V}_{m}^{\mathrm{D}}}
\frac{1}{c^2\sigma_{\Delta\tau,n}^{2}}
\left(
\frac{x-x_{T,n}}{r_{T,n}}
+
\frac{x-x_{R,n}}{r_{R,n}}
\right)
\nonumber\\
&\quad
\left(
\frac{z-z_{T,n}}{r_{T,n}}
+
\frac{z-z_{R,n}}{r_{R,n}}
\right),
\label{eq:Ixz_d_open}\\
I_{yz}^{\mathrm{D},m}
&=
\sum_{n\in\mathcal{V}_{m}^{\mathrm{D}}}
\frac{1}{c^2\sigma_{\Delta\tau,n}^{2}}
\left(
\frac{y-y_{T,n}}{r_{T,n}}
+
\frac{y-y_{R,n}}{r_{R,n}}
\right)
\nonumber\\
&\quad
\left(
\frac{z-z_{T,n}}{r_{T,n}}
+
\frac{z-z_{R,n}}{r_{R,n}}
\right).
\label{eq:Iyz_d_open}
\end{align}
\endgroup

\bibliographystyle{IEEEtran}
\bibliography{references}
\makeatletter\def\@IEEEBIOskipN{12pt}\makeatother
\begin{IEEEbiography}[{\includegraphics[width=1in,height=1.25in,clip,keepaspectratio]{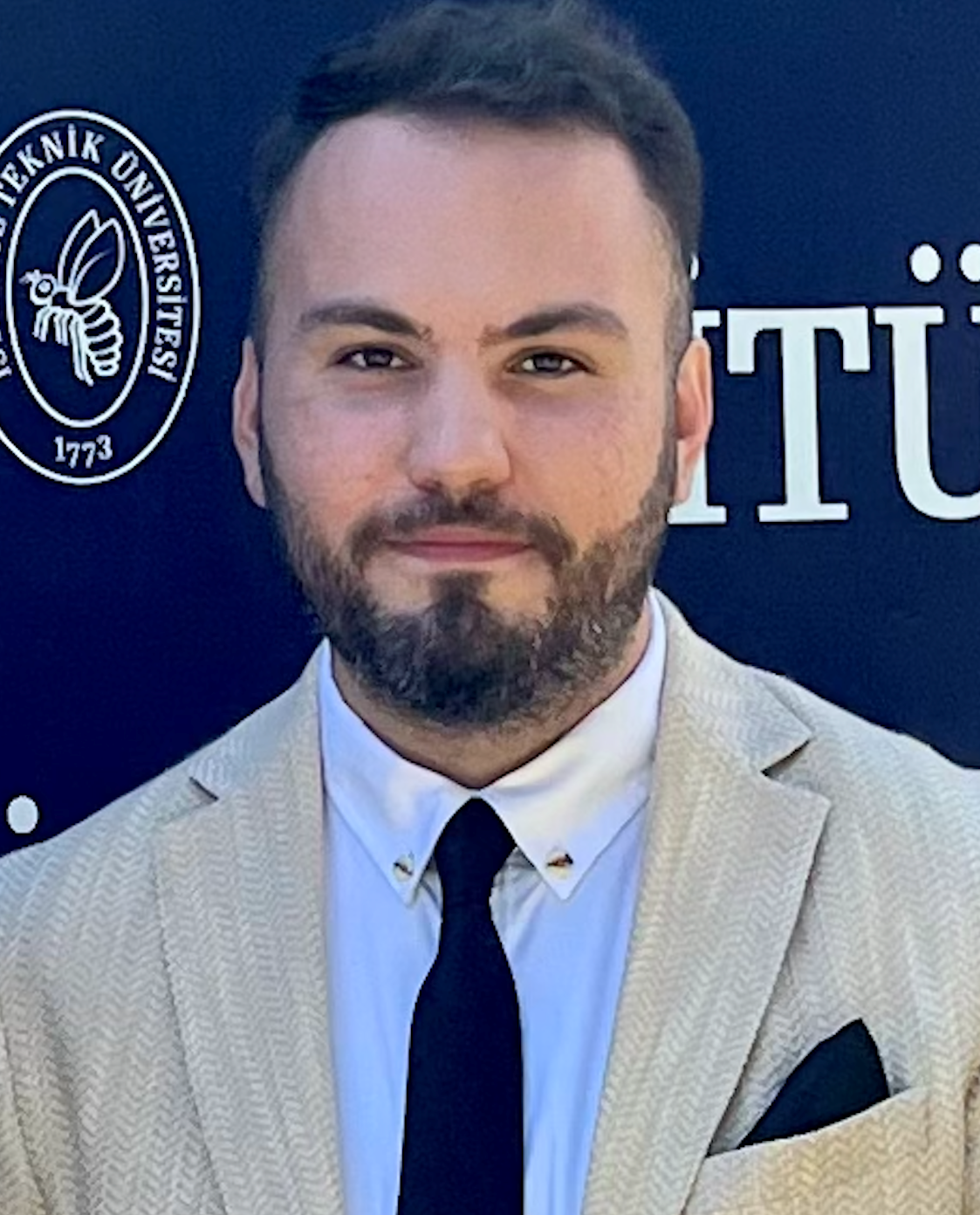}}]{Necati Kagan Erkek }
(Student Member, IEEE) received his B.Sc. degree in Electronics and Communication Engineering from Istanbul Technical University, where his bachelor’s thesis focused on communication channel modeling for unmanned aerial vehicles. He gained professional experience in wireless communication technologies in the defense industry at TÜBİTAK BİLGEM, Turkey. In 2024, he completed his M.Sc. degree in Telecommunication Engineering at Politecnico di Milano, with a thesis on AI-based origin identification of JPEG-based synthetic images. He is currently a Marie Skłodowska-Curie Actions (MSCA) fellow, funded by the European Union and UK Research and Innovation (UKRI). He is pursuing his Ph.D. at King’s College London under the supervision of Prof. Marco Di Renzo. In parallel, he is a Research and Innovation Engineer at InterDigital in London, where he works on AI/ML-based over-the-air computation and ISAC.
\end{IEEEbiography}

\begin{IEEEbiography}[{\includegraphics[width=1in,height=1.25in,clip,keepaspectratio]{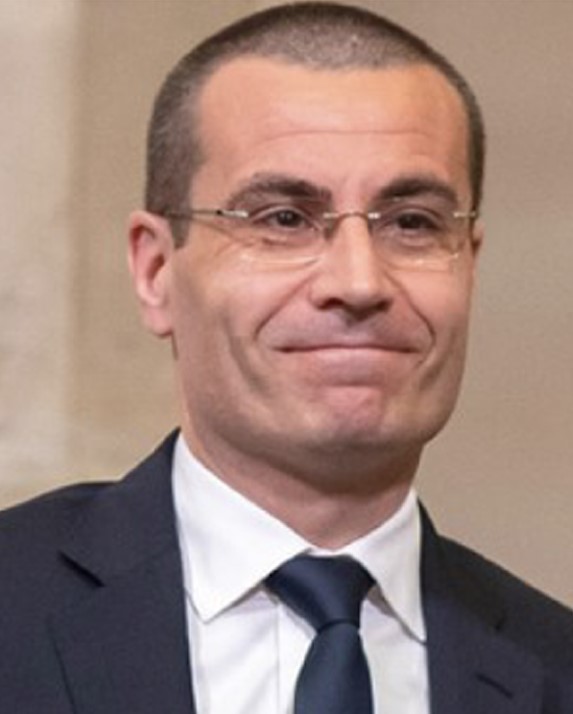}}]{Marco di Renzo }
(Fellow, IEEE) received the Laurea (cum laude) and Ph.D. degrees in electrical engineering from the University of L’Aquila, Italy, in 2003 and 2007, respectively, and the Habilitation ´a Diriger des Recherches (Doctor of Science) degree from Paris-Sud University (currently Paris Saclay University), Paris, France, in 2013. He was a France-Nokia Chair of Excellence in ICT with the University of Oulu, Finland, a Tan Chin Tuan Exchange Fellow in engineering with Nanyang Technological University, Singapore, a Fulbright Fellow with The City University of New York, USA, a Nokia Foundation Visiting Professor with Aalto University, Finland, and a Royal Academy of Engineering Distinguished Visiting Fellow with Queen’s University Belfast, U.K. Currently, he is the CNRS Research Director (a Professor) and the Head of the Intelligent Physical Communications Group with the Laboratory of Signals and Systems (L2S), CNRS, CentraleSup´elec, Paris-Saclay University, as well as a Chair Professor of telecommunications engineering with the Centre for Telecommunications Research, Department of Engineering, King’s College London, London, U.K. He is a fellow of IET, EURASIP, and AAIA; an Academician of AIIA; an Ordinary Member of European Academy of Sciences and Arts, Academia Europaea, and Italian Academy of Engineering and Technology; an Ambassador of European Association on Antennas and Propagation; and a Highly Cited Researcher. His recent research awards include the Michel Monpetit Prize conferred by the French Academy of Sciences, the IEEE Comm. Society Heinrich Hertz Award, and the IEEE Comm. Society Marconi Prize Paper Award in Wireless Comm. He served as the Editor-in-Chief for IEEE COMM. LETTERS from 2019 to 2023. His current main roles within IEEE Comm. Society include serving as a Voting Member of Fellow Evaluation Standing Committee, Chair of Publications Misconduct Ad Hoc Committee, and Director of Journals.
\end{IEEEbiography}

\begin{IEEEbiography}[{\includegraphics[width=1in,height=1.25in,clip,keepaspectratio]{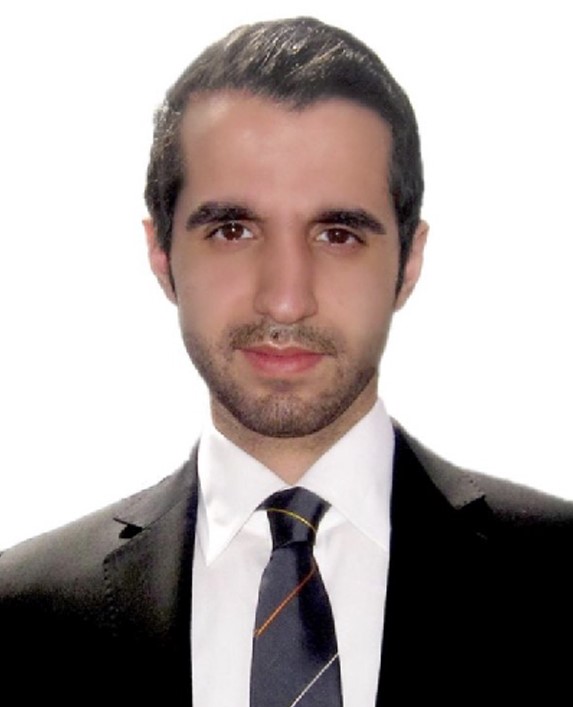}}]{Arman Shojaeifard }
(Senior Member, IEEE) received the B.Eng. degree in information systems engineering from Imperial College London in 2008 and the M.Sc. degree in signal processing and the Ph.D. degree in wireless communications from King’s College London in 2009 and 2012, respectively. He is currently a Senior Manager at InterDigital Inc., where he is responsible for the Radio Interface Research and Development at the Wireless Europe Laboratory (WLE). Prior to InterDigital, he was a Principal Researcher and the 3GPP RAN1 WG Delegate at British Telecommunications Plc. Prior to that, he held post doctoral research positions at University College London and The University of Manchester. He has experience in research, development, and standardization of wireless technologies. He serves as the Chair for the ETSI (ISG RIS).
\end{IEEEbiography}

\begin{IEEEbiography}[{\includegraphics[width=1in,height=1.25in,clip,keepaspectratio]{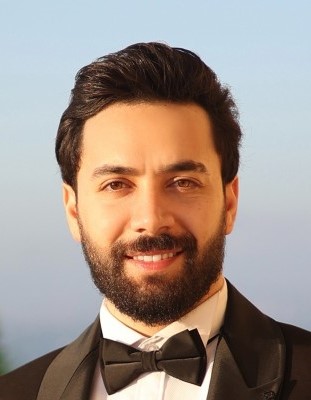}}]{Yasser Mestrah }(Senior Member, IEEE) received the Ph.D. degree in Electrical Engineering from IMT Nord Europe and Université de Reims Champagne-Ardenne (URCA), France, in 2019, under a full scholarship from Télécom Paris (2016–2019). From 2019 to 2021, he was awarded a competitive research scholarship to pursue postdoctoral research with INSA Rennes and ENENSYS Technologies, France, where he worked on advanced wireless systems and broadcast network virtualization. He then joined Mitsubishi Electric R\&D Centre Europe, Rennes, France (2021–2023), contributing to research on digital twins, private 5G networks, and machine learning-driven optimization. Subsequently, he joined InterDigital, London, U.K., where he is currently a R\&I Staff Engineer, contributing to RAN research and standardization within 3GPP and ETSI. He has filed more than 50 patents and authored several peer-reviewed publications in 5G/6G systems. He is actively involved in European research projects, academic collaborations and contributes to international standardization efforts. His research interests include AI/ML for wireless communications, information theory, integrated sensing and communication, and intelligent 6G network design.
\end{IEEEbiography}

\begin{IEEEbiography}[{\includegraphics[width=1in,height=1.25in,clip,keepaspectratio]{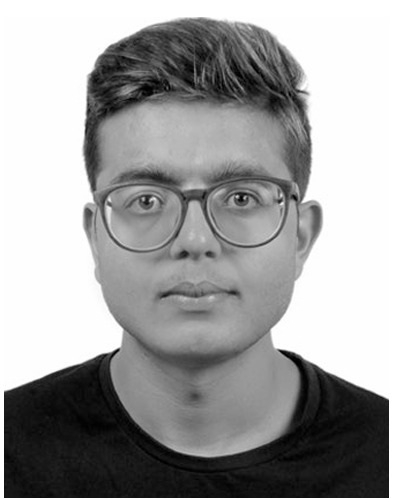}}]{ Remun Koirala }
(Member, IEEE) received his B.Sc. and M.Sc. degrees from Jacobs University, Bremen, Germany, in 2014 and 2016, respectively, and his Ph.D. degree jointly from the University of Rennes 1 and the University of Bologna in 2020. During his Ph.D., he focused on advanced topics in wireless communications, particularly in the area of millimeter-wave systems. His research interests include positioning and communication techniques in the context of millimeter-wave and beyond-5G systems. He is currently with InterDigital Inc., where he is involved in research and development of 6G physical layer technologies, including ISAC.
\end{IEEEbiography}

\begin{IEEEbiography}[{\includegraphics[width=1in,height=1.25in,clip,keepaspectratio]{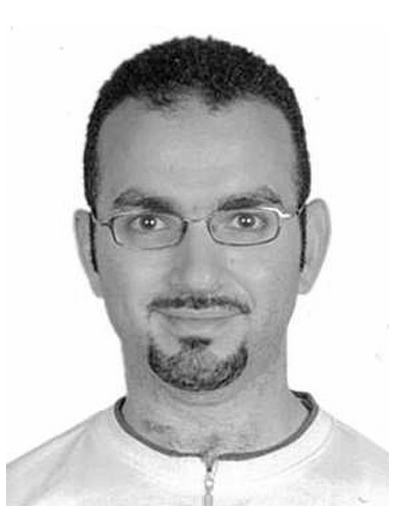}}]{ Mohammad Heggo }
(Member, IEEE) received the B.Sc. and M.Sc. degrees in electrical engineering from Ain Shams University, Cairo, Egypt, in 2006 and 2010, respectively, and the Ph.D. degree in electrical engineering from the University of Liverpool, Liverpool, U.K., in 2017. He is currently an Engineer II (Research Fellow) with the Research and Innovation Department, InterDigital Europe, London, U.K., working on advanced wireless and beyond-5G systems and contributing to international standardization and collaborative research. He previously held research positions at Imperial College London, the University of Manchester, and the University of Sheffield (with the Rolls-Royce University Technology Centre), focusing on RF sensing, autonomous systems, and wireless sensing in challenging environments. Earlier, he was a Research Assistant with A*STAR, Singapore. His research interests include advanced wireless systems, integrated sensing and communication (ISAC), electromagnetic-based non-intrusive monitoring, power line communications, IoT systems, and channel modeling. 
\end{IEEEbiography}

\begin{IEEEbiography}[{\includegraphics[width=1in,height=1.25in,clip,keepaspectratio]{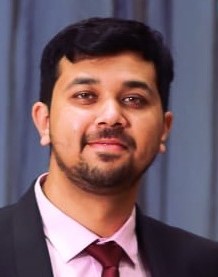}}]{Kunjan Shah }
(Member, IEEE) He is currently a Senior Staff Engineer at InterDigital, London, U.K. He received his B.E. degree in Electronics and Telecommunications from BVM College of Engineering, India, in 2012, and his M.Sc. degree in Electrical Engineering from Hochschule Darmstadt, Germany, in 2014. His research and professional interests include system-level simulation, wireless positioning, sensing technologies, and next-generation communication systems. 
\end{IEEEbiography}

\end{document}